\documentclass[reprint,superscriptaddress,amsmath,amssymb,aps,showkeys]{revtex4-2}
\usepackage{physics}
\usepackage{tabularx}
\usepackage{url}
\usepackage{svg}
\usepackage{xcolor,soul}
\usepackage{hyperref}
\usepackage{multirow}
\usepackage{graphicx}
\usepackage{dcolumn}
\usepackage{bm}
\begin{document}

\preprint{APS/123-QED}
\title{First-principles investigation of hydrogen-related reactions\\on (100)--(2$\times$1)$:$H diamond surfaces}

\author{Emerick Y. Guillaume}
\email[Corresponding author: ]{emerick.guillaume@uhasselt.be}
\affiliation{Namur Institute of Structured Matter (NISM), University of Namur, Rue de Bruxelles 61, 5000 Namur, Belgium}
\affiliation{Hasselt University, Institute for Materials Research (IMO-IMOMEC), 3590 Diepenbeek, Belgium}
\affiliation{IMOMEC, IMEC vzw, Wetenschapspark 1, 3590 Diepenbeek, Belgium}
\author{Danny E. P. Vanpoucke}
\affiliation{Hasselt University, Institute for Materials Research (IMO-IMOMEC), 3590 Diepenbeek, Belgium}
\affiliation{IMOMEC, IMEC vzw, Wetenschapspark 1, 3590 Diepenbeek, Belgium}
\author{Rozita Rouzbahani}
\affiliation{Hasselt University, Institute for Materials Research (IMO-IMOMEC), 3590 Diepenbeek, Belgium}
\affiliation{IMOMEC, IMEC vzw, Wetenschapspark 1, 3590 Diepenbeek, Belgium}
\author{Luna Pratali Maffei}
\affiliation{CRECK Modeling Lab, Politecnico di Milano, Piazza L. da Vinci, 32, 20133 Milano, Italy}
\author{Matteo Pelucchi}
\affiliation{CRECK Modeling Lab, Politecnico di Milano, Piazza L. da Vinci, 32, 20133 Milano, Italy}
\author{Yoann Olivier}
\affiliation{Namur Institute of Structured Matter (NISM), University of Namur, Rue de Bruxelles 61, 5000 Namur, Belgium}
\author{Luc Henrard}
\affiliation{Namur Institute of Structured Matter (NISM), University of Namur, Rue de Bruxelles 61, 5000 Namur, Belgium}
\author{Ken Haenen}
\affiliation{Hasselt University, Institute for Materials Research (IMO-IMOMEC), 3590 Diepenbeek, Belgium}
\affiliation{IMOMEC, IMEC vzw, Wetenschapspark 1, 3590 Diepenbeek, Belgium}

\date{\today}

\begin{abstract}
Hydrogen radical attacks and subsequent hydrogen migrations are considered to play an important role in the atomic-scale mechanisms of diamond chemical vapour deposition growth. We perform a comprehensive analysis of the reactions involving H-radical and vacancies on H-passivated diamond surfaces exposed to hydrogen radical-rich atmosphere. By means of first principles calculations---density functional theory and climbing image nudged elastic band method---transition states related to these mechanisms are identified and characterised. In addition, accurate reaction rates are computed using variational transition state theory. Together, these methods provide---for a broad range of temperatures and hydrogen radical concentrations---a picture of the relative likelihood of the migration or radical attack processes, along with a statistical description of the hydrogen coverage fraction of the (100) H-passivated surface, refining earlier results via a more thorough analysis of the processes at stake. Additionally, the migration of H-vacancy is shown to be anisotropic, and occurring preferentially across the dimer rows of the reconstructed surface. The approach used in this work can be generalised to other crystallographic orientations of diamond surfaces or other semiconductors.
\end{abstract}


\maketitle


\section{\label{sec:level1}Introduction}

For decades, studies have shown that hydrogen plays a central role during diamond growth by means of chemical vapour deposition (CVD).\cite{Fedoseev1984} Numerous authors posited that its importance lies in etching away graphitic forms of carbon, leaving the diamond intact.\cite{Frenklach1989} However, this description of hydrogen effects on diamond growth remains incomplete as it disregards its possible importance as a surface site activator for both fully passivated and bare surface dimers. This uncertainty, related to the role of hydrogen radical attack, also originates from the lack of consensus among early studies showing either a large barrier or no barrier at all for the desorption of hydrogen via radical attack of H. It was either claimed to be the most difficult reaction to overcome at CVD conditions, making it the limiting reaction of the diamond growth,\cite{Frenklach1988,Deak1991,Harris1990,Frenklach1991} or suggested to occur spontaneously by other authors.\cite{Valone1990} According to more recent work,\cite{Cheesman2008,Oberg2021} neither assumptions are correct. Indeed, considering a (100)--(2$\times$1)$:$H diamond surface (known to reconstruct along dimer rows \cite{Brenner1990,Hamza1990,Tsuno1991}), these studies showed that energy barriers encountered during the carbon insertion into the surface dimers is substantially larger than that of a hydrogen reaction activating the surface. In contrast, the work presented here focuses more on the conditions that trigger the growth rather than on the growth itself. As a consequence, our investigations of the migrations are restricted to the movement of vacancies on a surface with no terrace-like features since movement near terraces is rather a matter of further growth from a nucleated seed.

Building on the most recent studies \cite{Cheesman2008,Oberg2021} which investigate reaction paths in terms of energy barriers, our work focuses on the calculation of the reaction rates and includes the ab-initio calculation of the exponential prefactors, which allows for an even broader picture of the atomic-scale mechanisms taking place on diamond surfaces. As a matter of fact, this approach is more appropriate to derive statistical properties and to further investigate the growth by means of macro-scale modelling, for instance by implementing the findings in detailed and semi-detailed chemistry models available for CVD processes\cite{Frenklach1992,Serse2023}.

The CVD growth of diamond relies on complex interactions between diamond surface and gas phase species in the chamber. In the context of lab-grown diamond, the gas phase plays a crucial role (concentration and type of different molecules and radicals). Since the present study focuses on hydrogen-related reaction on diamond surface and not on the kinetics governing the gas phase, we initially assume a representative molar concentration of 1\% of hydrogen radical $\mathrm{H}^{\cdot}$ and 99\% of molecular hydrogen $\mathrm{H}_2$.\cite{Harris1990c} It allows to accommodate for the presence of gas phase radical and molecules, whereupon we discuss how the radical-to-molecule ratio influences macroscopic properties such as the hydrogen surface coverage of the diamond surface. This specific property has already been investigated elsewhere,\cite{May2010} but to the best of our knowledge, most studies (\textit{e.g.} \cite{Netto2005,May2007}) investigating the macroscopic effect of hydrogen-related reactions on diamond (100) surfaces directly or indirectly refers to early measurement of the reaction rate for hydrogen ad/desorption.\cite{Krasnoperov1993} Given that the sample used in this experiment seems to be an abrasive or cutting disk for mechanic purposes, it is difficult to conclude on the reaction rate coefficient value of a single particular reaction due to low sample purity and uncertainties about its precise crystalline structure.

This work aims at providing a broad view of the mechanisms involved in these interactions through accurate, atomic-scale descriptions of the energy and geometrical configurations of the transition states (TS) of the different chemical reactions driving the early stages of the diamond growth. Using state-of-the-art numerical methods---density functional theory (DFT) \cite{Hohenberg1964,Kohn1965} along with the (climbing image) nudged elastic band, (c)NEB, \cite{Henkelman2000}---we compute geometries, energies and vibrational spectrum of the initial, final and transition states of gas-surface interactions. During these reactions, the distances between molecules (or radicals) and the surface vary from about 1 to 5 \AA. To characterise a given reaction, it is essential to describe the intermediate configurations with high accuracy. Therefore, we also investigate the role of van der Waals (vdW) interactions on the energy barriers by considering different long-range energy correction schemes to the DFT calculations.  

Our analysis focuses on hydrogen desorption via hydrogen radical attack, hydrogen radical recombination with surface radical sites and migration of H-vacancies on (100) H-passivated diamond surfaces. We assume the surfaces to be exposed to a gas mixture of molecular (H$_2$) and radical (H$^{\cdot}$) hydrogen generated by means of a microwave plasma or a hot filament. The interactions of methyl radicals and other gas-phase carbonaceous species is considered to be outside of the scope of the current work, yet being an interesting subject for future research through an extension of the present methodology to these species. 

This article is structured as follows; in section \ref{sec:comp_methods} the numerical methods and the system modelling are introduced and the influence of different vdW dispersion corrections on the calculated energy barriers is presented in Table \ref{tab:vdw+kpts}. This is followed by an in-depth investigation of the hydrogen adsorption and desorption processes in sections \ref{subsec:H_rad_1st}, \ref{subsec:H_rad_2nd} and \ref{subsec:H_recomb}, with an emphasis on the geometric description of the TSs. Using the reaction rate coefficients of the gas-surface reactions listed in Table \ref{tab:reaction_1-6}, the surface coverage of activated sites is discussed in section \ref{subsec:H_surf_cov}. The possibility for hydrogen migration and diffusion either between radical sites on a H-passivated surface or between radical sites on a clean surface are then investigated in sections \ref{subsec:H_mig} and \ref{subsec:H_vac_mig}, and an overview of the reaction rate coefficients (forward and backward direction) for these surface reactions is provided in Table \ref{tab:E_Mob_H}. Finally, accurate reaction rate coefficients for the different mechanisms are calculated, allowing to determine the equilibrium outcome of the competing phenomena. Such an assessment is crucial to determine the effective best conditions to grow CVD diamond upon implementation of our findings in existing or future kinetic models.

\section{\label{sec:comp_methods}Materials modelling and computational methods}
To model a two-dimensional infinite diamond surface, we first account for the 2$\times$1 reconstruction along dimer rows. However, depending on the number of layers of the slab, dimer rows on both surfaces may be oriented parallel or perpendicular to one another. We therefore consider an even repetition of the surface along both $x$ and $y$ directions and we represent the diamond surface using a periodic 4$\times$4$\times$11 slab model containing 176 C atoms and 32 H atoms passivating the surface (Fig. \ref{fig:slab}). In the Supplementary Information, we show that eleven carbon layers and a vacuum distance of 10 \AA$ $ between periodic repetitions of the slab along the z-direction is sufficient to achieve convergence of the surface energy. 

As this work focuses on the reactions happening on the diamond surface, only the hydrogenation states of the surface dimers are investigated: a dimer can be found in its fully passivated ($\mathrm{RH}_2$), half-passivated ($\mathrm{R^{\vdot}H}$) or non-passivated state ($\mathrm{R}$).

\begin{figure}[!tbh]
    \begin{center}
        \includegraphics[scale=0.40]{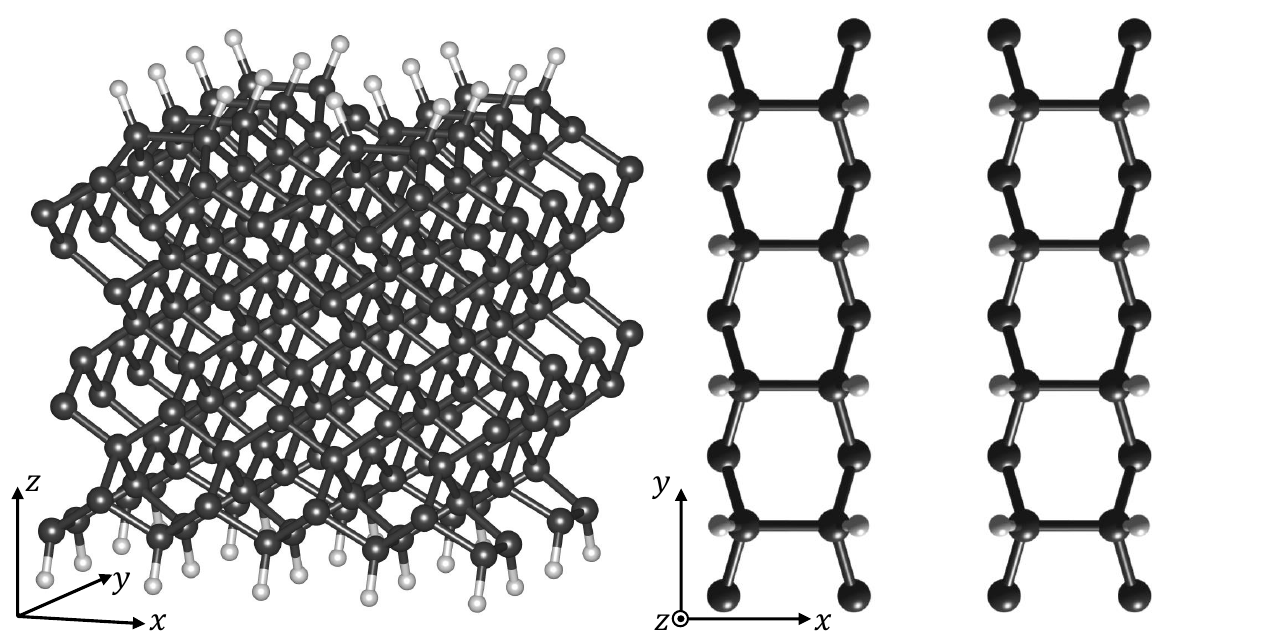}
        \caption{Left: An entire (100)-orientated H-passivated slab, as used in the calculations. Right: Top view of the slab. For the sake of visualisation, only the top two C-layers and atoms above are depicted. Black and white atoms are carbon and hydrogen atoms, respectively. This convention will be used throughout this work.}
        \label{fig:slab}
    \end{center}
\end{figure}

To calculate the reaction rate coefficients using the transition state theory (TST), the first step of our methodology is to characterise the energy and geometry of the system evolving along the minimum energy path (MEP) of each reaction. There exist several methods (\textit{e.g.} dimer method or molecular dynamics) that accurately determine the MEP, of which the cNEB---an extension of the NEB---is used in this work. The NEB method \cite{Jonsson1998} considers multiple intermediate configurations (list of atom positions, also known as \textit{images}) between two stable states. To prevent the images from relaxing towards the nearest stable state, they are connected through fictional springs thus forming what can be seen as an \textit{elastic rubber band}. The algorithm allows the images to relax according to the potential gradient as well as to the effect of the fictitious springs that connect successive images. This allows the set of images to move towards the MEP. cNEB extend the usual NEB method by ensuring that one of the images moves to the maximum energy point along the MEP (provided the existence of an energy barrier), \textit{i.e.} to the saddle point in between initial and final states.\cite{Henkelman2000,Henkelman2000b}. Both the NEB and cNEB are local, not global, optimisers. As such, they can only find the nearest MEP to the initial reaction path that is fed to the algorithm. Specific details about the initialisation of these methods is provided later. cNEB forces and energies are obtained through DFT calculations, carried out using the Vienna Ab initio Simulation Package (VASP) \cite{Kresse1993_VASP,Kresse1996_VASP,Kresse1996b_VASP,Kresse1999_VASP} including the VASP Transition State Tools plugin \cite{VTST} to allow for cNEB calculations. Using DFT for the cNEB calculations comes with limitation on the reactions that are considered in this work: the spin should not be allowed to change between reactants and products. As such, different spin configurations of the reactants lead to multiple individual paths which products are different spin configurations of the same system. In this work we focus on the path leading to the most stable products (singlet states). However, for the sake of completion, we also provide the reaction rate coefficient of the reactions leading to triplet state in the SI.

Since we are dealing with gas-surface interactions---where radicals or molecules approach the surface to create defects and/or break bonds---it is necessary to consider vdW dispersion corrections in our DFT calculations as the distance between gas and surface falls into the range of these interaction at different points along the reaction path. Accordingly, we briefly discuss the importance of considering a vdW dispersion correction scheme and show, in section \ref{subsec:H_rad_1st}, the common features to different implementation available in VASP (DFT-D2 \cite{Grimme2006}, DFT-D3 \cite{Grimme2011}, Tkatchenko-Scheffler (Tk.-Sc.) \cite{Tkatchenko2009} and density-dependent dispersion correction (DDsC)\cite{Steinmann2011,Steinmann2011b}).

In the context of reaction rate coefficient calculation, it is useful to distinguish loose and tight transition states (noted tTS and lTS). In the following, TST will be applied to reactions showing a well-defined saddle point on their potential energy surface, whereas the variational form of TST (VTST) will be applied to reactions which do not. Although the distinction between lTS and tTS depends on many features of the given TS, in this specific investigation of reactions on surfaces, all lTS relate to the barrierless reactions and all tTS relate to reactions showing an energy barrier.

\subsection{\label{subsec:TST}Tight transition states (tTS)}
Characterisation of a tTS requires first to determine the MEP using the cNEB method. This provides both the energy and the geometry of the TS, tightened to the highest energy point along the reaction path. To compute the reaction rate coefficient, TST requires the calculation of the partition functions.\cite{Atkins2014} Briefly, these reaction rate coefficients $k_n$ can be computed as 
\begin{equation}
    k_n=\kappa\left(T,E_{n}^{TS}\right)\frac{k_BT}{h}\frac{Q_{n}^{*TS}}{Q_{n}^{I}},
    \label{eq:reaction_rate_prim}
\end{equation}

where $\kappa$ is a transmission coefficient to account for tunnelling effects (in this work we consider the implementation of Skodje and Truhlar \cite{Skodje1981}) and $Q_{n}^{I}$ is the partition function of the reactant. The latter is often expressed as the product of the translational ($q_t$), rotational ($q_r$), vibrational ($q_v$) and electronic ($q_{el}$) partition functions:
\begin{equation}
    \begin{aligned}
        Q=&q_{t}q_{r}q_{v}q_{el}\\
        \text{ $ $ with $ $ }&q_{t}=\left(\frac{2\pi m_gk_BT}{h^2}\right)^{3/2}\\
        &q_{r}=\left(\frac{1}{\sigma}\right)\sum_{J=0}^{\infty}\left(2J+1\right)\exp\left(\frac{-J\left(J+1\right)h^2}{8\pi^2I_gk_BT}\right)\\
        &q_{v}=\exp\left(-\frac{E_{ZPE}}{k_BT}\right)\prod_j^N\frac{1}{1-\exp\left(\frac{-h\nu_j}{k_BT}\right)}\\
        &q_{el}=\sum_lg_l\exp\left(\frac{-E_l}{k_BT}\right)
    \end{aligned}
    \label{eq:partition_function}
\end{equation}

Here $m_g$ and $I_g$ are the mass and moment of inertia of the gas phase reagent, $\sigma$ is the symmetry number ($\sigma=2$ for H$_2$) and $\nu_j$ are the eigenvalues (\textit{i.e.} vibrational frequencies) of the dynamical matrix. $k_B$ is the Boltzmann constant, $h$ is the planck constant and $T$ is the temperature. Note that the formula for the rotational partition function $q_r$ only applies for a diatomic molecule (H$_2$ in this work). $E_{ZPE}$ is the zero-point energy, while $E_l$ and $g_l$ are the electronic energies and their respective degeneracy, respectively.


The $Q_n^{*TS}$ term in eq. (\ref{eq:reaction_rate_prim}) is the quasi-partition function of the TS: the vibrational partition function of the TS is computed as shown in eq. (\ref{eq:partition_function}) considering all $\nu_j$ except the decomposition frequency of the TS into the products (\textit{i.e.}, the only imaginary eigenvalue of the dynamical matrix). Due to the periodic boundary conditions applied to the slabs, it prevents them from rotating, therefore their rotational partition functions equal one. 

Finally, reaction rate coefficients $k_n$ can be expressed as:
\begin{equation}
    \begin{aligned}
        k_n&=\kappa\left(T,E_{n}^{TS}\right)\frac{k_BT}{h}\frac{q_{v,n}^{*TS}}{q_{t,n}^{I}q_{r,n}^{I}q_{v,n}^{I}}\times\frac{q_{el,n}^{TS}}{q_{el,n}^{I}}\\
        &=A_n\times\exp\left(\frac{-E_n^{TS}}{k_BT}\right)
    \end{aligned}
    \label{eq:reaction_rate_fin}
\end{equation}

where $A_n$ refers to the exponential prefactor or collision frequency factor, that encompasses all terms but the exponential. Depending on the reaction and its direction, $A_n$ (and consequently $k_n$) can have different units. Reaction rate coefficients $k_n$ (and prefactor $A_n$) whose reactants include a gas phase radical or molecule are expressed in units of cm$^3\vdot$mol$^{-1}\vdot$s$^{-1}$, whereas those which do not involve any gas phase reactants are expressed in s$^{-1}$. 

To remedy the complication of comparing values expressed in different units, we multiply the reaction rate coefficients with the density of the associated gas-phase reactant to provide the reader with an \textit{effective} reaction rate coefficient $r_n$ considering usual CVD conditions.\cite{Harris1990c,Rouzbahani2021} 

\subsection{\label{subsec:VTST}Loose transition states (lTS)}
To compute the reaction rate coefficient of a lTS, we consider the TST in its variational form (VTST). Practically, VTST demands a sampling of the systems along the MEP to compute the flux along this path considering eqs. (\ref{eq:reaction_rate_prim}--\ref{eq:reaction_rate_fin}). In practice, eq. (\ref{eq:reaction_rate_fin}) is computed assuming that each point $\bar{s}$ along the reaction path is the TS ($\bar{\cdot}$ denotes a $3N$-dimensional vector where $N$ is the total number of atoms in the slab and gas phase). VTST posits that the lowest of these fluxes $k\left(T,\bar{s}\right)$ is associated to the actual TS, at the temperature considered in eq. (\ref{eq:min_k_s}). 
In other words, the TS is loose and VTST determines, for each temperature, where the kinetic bottleneck---which is the limiting step of the reaction---lies along the reaction path. 
\begin{equation}
    \begin{aligned}
        k\left(T\right)&=\min_{\bar{s}\in MEP}\left\{k\left(T,\bar{s}\right)\right\}\\&=\min_{\bar{s}\in MEP}\left\{A_0\left(T,\bar{s}\right)\times\exp\left(-\frac{E^{\bar{s}}}{k_BT}\right)\right\}.
    \end{aligned}
    \label{eq:min_k_s}
\end{equation}

It is important to ensure that the sampling is dense enough to locate the points along the MEP that minimise the reaction rate coefficient for a specific range of temperature. The NEB method proves useful, either through an increase of the total number of images or via subsequent NEB runs between images along the MEP to obtain decent sampling of the relevant sampling area.

\subsection{\label{subsec:Num_settings}Numerical settings}
In addition to the vdW dispersion correction to the PBE functional, our calculations are spin-polarised to accommodate the radical nature of the atoms and molecules. A 4$\times$4$\times$1 $\Gamma$-centred sampling of the Brillouin zone (BZ)---resulting in 10 k-points in the irreducible BZ---is initially considered for the first hydrogen radical attack. As demonstrated in the following, results show that a $\Gamma$-only sampling of the BZ provides the best trade-off between accuracy and computational efficiency. The cut-off energy for the plane-wave basis is set to 650 eV. The self-consistent field criterion for convergence is fixed to $10^{-6}$ eV and the geometric optimisation criterion on forces is 0.02 eV/\AA. The bottom 16 H atoms and the lowest 16 C atoms (\textit{i.e.} the bottom layer of carbon and their passivating hydrogen, see Fig. \ref{fig:slab}) are kept frozen during the cNEB calculations to avoid the whole slab to move inside the cell. Furthermore, this allows us to account for a bulk-like structure in the depth of the surface. 

Vibrational spectrum calculations are carried out using finite difference method as implemented in VASP, using a displacement of 0.02 $\text{\AA}$ and a $\Gamma$-only sampling of the BZ. The dynamical matrix is then diagonalised using the HIVE3 software\cite{Vanpoucke2020}. All degrees of freedom are considered during these calculations (no atoms are kept frozen) and electronic convergence criteria is also $10^{-6}$ eV.

For the calculation of the unimolecular dissociation reaction rate coefficients (\textit{i.e.} reactions 1 and 3 in Table \ref{tab:reaction_1-6}), the atom to be dissociated is pulled away from the surface with increments of 0.05 \AA. For each position, a relaxation of the whole system (except the height of the atom being pulled and the bottom layers of carbon and hydrogen of the slab) is performed prior to the vibrational spectrum calculation. As highlighted in the Supplementary Information, the relevant sampling area for these reactions is located between 1.75 and 2.5 \AA $ $ away from the surface for temperatures around 1200 K. This procedure not only provides the reaction rate coefficient of the unimolecular dissociation, but also of its reverse reaction, \textit{i.e.} the bimolecular recombination.

The reaction path for the (c)NEB calculations is initialised---in case of a radical attack (\textit{i.e.} reactions 2 and 4)---as follows: 
\begin{enumerate}
    \item For both initial and final states, incoming/returning radicals or molecules are kept frozen at a distance of about 5 \AA$ $ above the surface to avoid interaction between the surface and the molecule prior to the reaction, while the atoms of the slab are able to relax.
    \item To ensure the barrier is physically relevant, the initial guess set of images must show that a radical comes close to the surface, then moves away as a different radical or molecule. To do so, an intermediate state is chosen, presumed to be close to the actual TS and for which the radical is very close to the surface. 
    \item Two interpolations are performed: a) between initial and intermediate states and b) between intermediate and final states.
\end{enumerate}

This initial guess constitutes somewhat an educated guess of the actual reaction path. For the sake of discussion, we only take interest in the values at $T=1200$ K in the main text. However, to capture the complexity of the temperature dependence, we provide a simple description of the reaction rate coefficients through a fit of the reaction rate coefficients to a modified Arrhenius equation:
\begin{equation}
    k_n(T)=a_nT^{b_n}\exp\left(\frac{-c_n}{k_BT}\right).
\end{equation}

The three parameters $a$, $b$, $c$ of each reaction are provided, along with details about the quality of the fit, in the Supplementary Information to allow for possible use in kinetic modelling covering a broad range of temperatures.

\section{\label{sec:level3}Results and discussions}
\subsection{\label{subsec:H_rad_1st}Hydrogen radical attacks : first H-desorption}
\begin{figure}[!tbh]
    \begin{center}
        \begin{center}
            \includegraphics[scale=0.22]{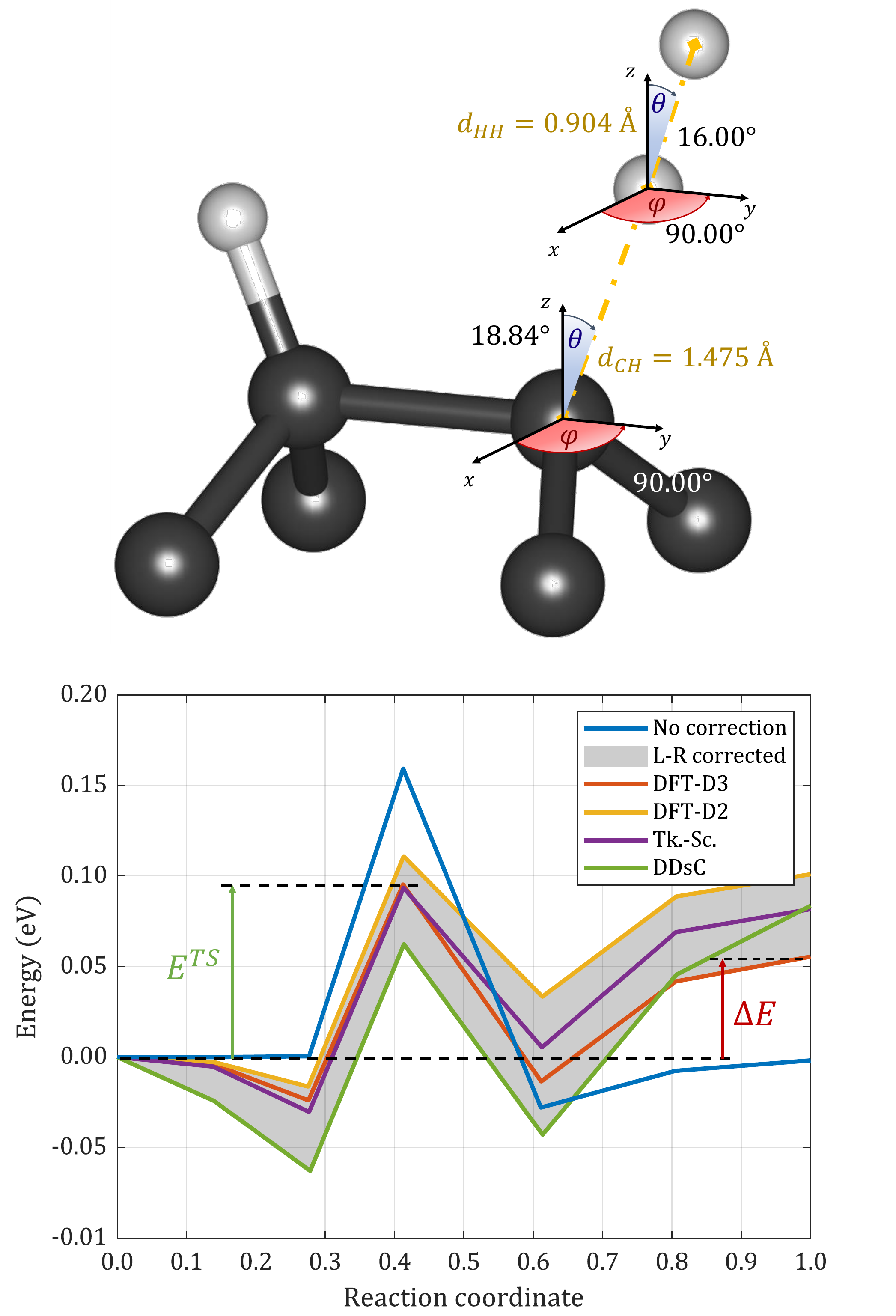}
        \end{center}
        \caption{Hydrogen radical attack on a H-passivated dimer (reaction 2 in Table \ref{tab:reaction_1-6}). Top: Geometry of the TS considering a 4$\times$4$\times$1 k-point sampling and DFT-D3 with B.-J. damping. $d_{CH}$ and $d_{HH}$ are the distances between the passivating hydrogen and surface carbon atom or the incoming radical hydrogen, respectively. $\theta$ and $\varphi$ are the spherical coordinates centred on each atom. The three atoms involved in the dimer breaking/forming are located in the same $yOz$ plane as the dimer and CH bonds of the passivated dimer. These three atoms form an almost linear configuration with a small deviation, probably due to the surface hydrogen atoms belonging to adjacent dimers in the $xOy$ plane, which are not depicted here. Bottom: Energy of the images for different vdW dispersion correction schemes. The grey zone highlights the energy range obtained with different vdW dispersion corrections. x-axis shows the reaction coordinate, that describes the progress of the reaction between reactants ($0$) and products ($1$).}
        \label{fig:reaction_2}
    \end{center}
\end{figure}

In terms of CPU time and accuracy, DFT-D3 is, to the best of our knowledge, the most efficient and robust vdW dispersion correction scheme for large systems as those investigated in this work \cite{Hermann2017}. As shown in Fig. \ref{fig:reaction_2}, calculations without vdW dispersion corrections (solid blue line) leads to a noticeable discrepancy with respect to DFT-D3 calculation (solid red line). To ensure that it is not a DFT-D3 specific shortcoming, and to investigate the influence of different vdW dispersion implementation, we compare the results to three other correction schemes: DFT-D2, Tkatchenko-Scheffler (Tk.-Sc.) and Density-Dependent energy Correction (DDsC). These additional energy profiles are also reported in Fig. \ref{fig:reaction_2}. 


$E^{TS}$ and $\Delta E$ values are defined as the differences between the energies of the TS and the final state with respect to the energy of the initial state (Fig. \ref{fig:reaction_2}), respectively.

\begin{table}[!tbh]
    \centering
    \caption{Summary of the TS energies and energy differences of the hydrogen radical attack on a H-passivated dimer process according to the set of parameters of each calculation. Energies are expressed in eV. Perdew-Burke-Ernzerhof (PBE) \cite{Perdew1996} is the semilocal exchange-correlation functional (XC) considered for this work. The long-range interaction corrections considered are the DFT-D2 \cite{Grimme2006}, DFT-D3 \cite{Grimme2011}), Tk.-Sc. \cite{Tkatchenko2009} and DDsC \cite{Steinmann2011,Steinmann2011b} methods.}
    \label{tab:vdw+kpts}
        \begin{tabularx}{0.48\textwidth}{c@{\extracolsep{\fill}}c@{\extracolsep{\fill}}c@{\extracolsep{\fill}} c@{\extracolsep{\fill}} c@{\extracolsep{\fill}} c}
            \hline
            \multicolumn{6}{|c|}{Hydrogen desorption}\\
            \hline
            \hline
            & correction scheme & & $\Gamma$ only & $4\times4\times1$ \\
            \hline
            \hline
            & \multirow{2}{*}{None} & 
            $\Delta E$ & -0.006 & -0.002\\
            & & $E^{TS}$   &  0.154 &  0.159\\  
            \hline
            & \multirow{2}{*}{DFT-D3} &
            $\Delta E$ &  0.052 &  0.056\\
            & & $E^{TS}$   &  0.091 &  0.095\\
            \hline
            & \multirow{2}{*}{DFT-D2} &
            $\Delta E$ & - &  0.101\\
            & & $E^{TS}$   & - &  0.111\\
            \hline
            & \multirow{2}{*}{Tk.-Sc.} &
            $\Delta E$ & - &  0.082 \\
            & & $E^{TS}$   & - &  0.093 \\
            \hline
            & \multirow{2}{*}{DDsC} & 
            $\Delta E$ & - &  0.084& \\
            & & $E^{TS}$   & - &  0.062& \\
            \hline
        \end{tabularx}
\end{table}

The energy profiles along the reaction coordinate reveal, for all vdW dispersion-corrected calculations, one maximum (the energy barrier corresponding to the TS), a positive $\Delta E$ and two minima. We suspect the positive $\Delta E$ arise from the interaction of the radical site with a hydrogen atom bound to a neighbouring dimer whereas the minima arise from the consideration of vdW interactions: the total energy of the system decreases and gets more stable as the H is approaching the surface, reaching a minimum when the H atom is about 2.5 \AA $ $ above the surface hydrogen to be etched. The same observation applies to the exit minimum, which is reached when the H$_2$ is about 2.9 \AA $ $ above the non-passivated carbon atom; as such the two minima represent vdW bonded configurations, sometimes referred to as vdW complexes. These complexes do not stabilise well, especially at temperature usually considered in CVD experiments. As such we chose the reactants and products as start and end points of the reactions.


Considering a full slab and accounting for vdW dispersion interactions through different methods leads to noticeably lower energy barriers than those found in the literature. Cheesman \textit{et al.} \cite{Cheesman2008}, Yiming \textit{et al.} \cite{Yiming2014} and Oberg \textit{et al.} \cite{Oberg2021} reported energy barriers of 0.27, 0.12 and 0.19 eV respectively, higher than the 0.095 eV value we report in the present work for DFT-D3 and 4$\times$4$\times$1 k-point sampling of the BZ (\textit{cf.} Table \ref{tab:vdw+kpts}). This discrepancy arises from the fact that, on the one hand, in ref. \cite{Cheesman2008}, only a fraction of the slab was computed at the DFT level. This combination of B3LYP with molecular mechanics seems to lead to some artificial intermediate steps in the growth, as discussed in ref. \cite{Oberg2021}. On the other hand, Yiming and Oberg (and respective co-workers) did not account for vdW dispersion interactions \cite{Yiming2014,Oberg2021} and considered a 3$\times$2$\times$1 and 1$\times$1$\times$1 sampling of the BZ respectively. The importance of vdW dispersion correction has been discussed earlier and explain the discrepancy, but k-point sampling of the BZ is found to have little impact on the energy (\textit{cf.} Table \ref{tab:vdw+kpts}). Based on this evidence and considering the prohibitive computational cost of vibrational calculation of a 4$\times$4$\times$1 sampling of the BZ, all calculations are performed using a 1$\times$1$\times$1 sampling.

Calculation of the vibrational spectra of the initial, final and transition states results in forward and backward effective reaction rate coefficients $r$ of $6.143\times10^{5}$ s$^{-1}$ and $4.252\times10^{5}$ s$^{-1}$, respectively (specific details about the reaction rate coefficients can be found in Table \ref{tab:reaction_1-6}).

\subsection{\label{subsec:H_rad_2nd}Hydrogen radical attacks : second H-desorption}
The half-passivated dimer resulting from the first radical attack can also be subject to an additional radical attack. Given the methods employed in this work, spin-configuration must be initialised to ensure that we only focus on the production of double-bonded dimers (singlet state). The reaction leading to a double-bonded, non-passivated dimer occurs through a lTS, as suggested by the energy profile shown in Fig. \ref{fig:Reaction_3}. Forward and backward reaction rate coefficients are computed following section \ref{subsec:VTST}, which results in values of $3.364\times10^{6}$  s$^{-1}$ and $2.836\times10^{2}$ s$^{-1}$, respectively. For the sake of completion, we also investigate the reaction rate coefficient of the production of non-passivated single-bonded dimers in the SI.

Making a clear distinction between first and second hydrogen desorption provides an improved picture in comparison with the existing literature.\cite{Skokov1994b} As a result, the surface coverage fractions of the different types of dimers are expected to differ from previously published results, as discussed later in section \ref{subsec:H_surf_cov}. 

\subsection{\label{subsec:H_recomb}Unimolecular dissociation and bimolecular recombinations}
There exist other ways to adsorb or desorb hydrogen atoms. In this section, we focus on the unimolecular dissociation and its reverse mechanism, the bimolecular recombination (reactions 1 and 3). The latter is usually more interesting as its reaction rate coefficient is expected to be orders of magnitude larger than the one of the former, due to the energy difference favouring the recombined form.

\begin{figure*}[!htbp]
    \begin{minipage}[t]{\columnwidth}
    \begin{center}
        \includegraphics[scale=0.22]{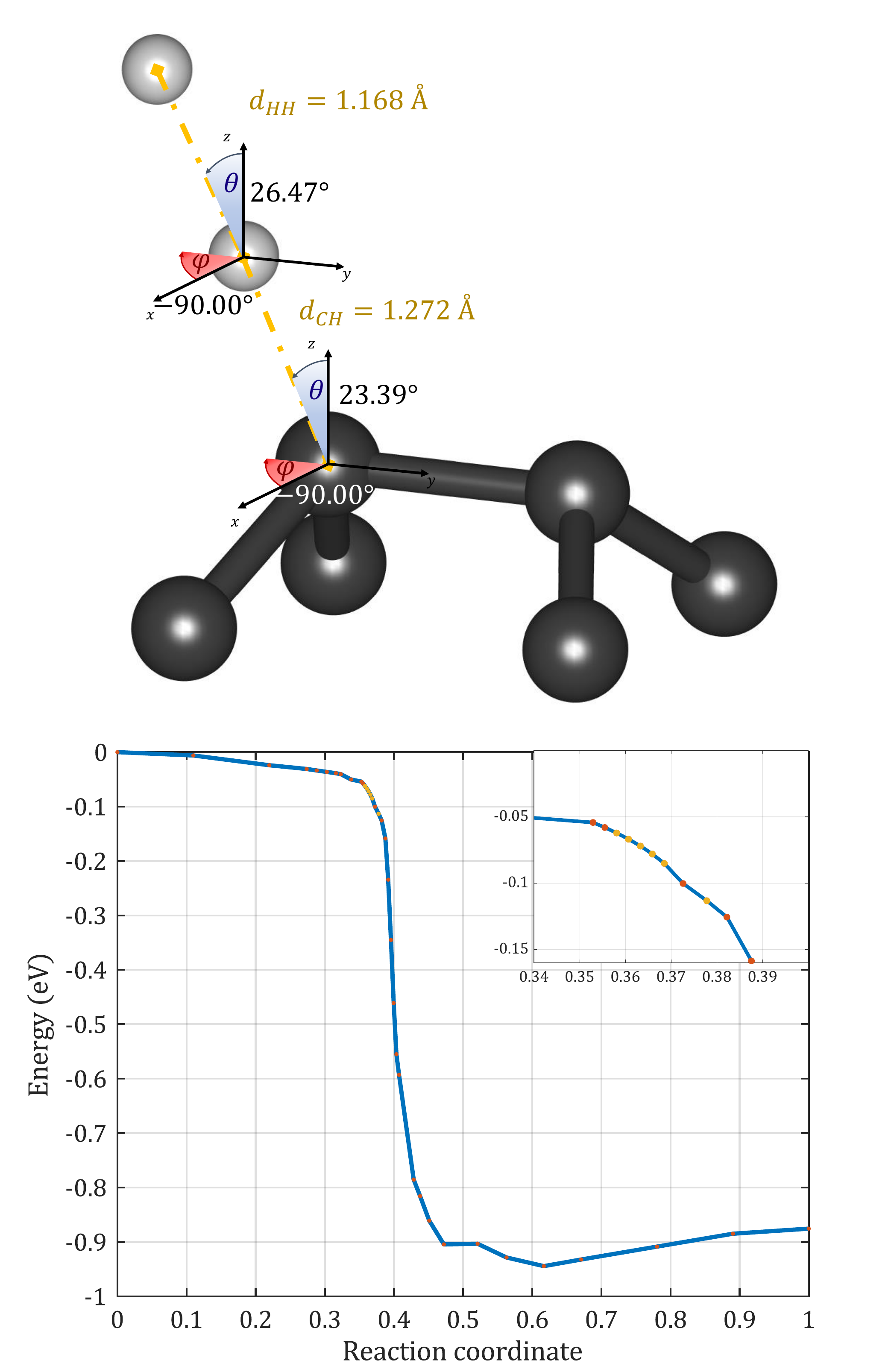}
    \end{center}
    \caption{Hydrogen radical attack on a half-passivated dimer (reaction 5 in Table \ref{tab:reaction_1-6}): $\mathrm{R^{\vdot}H}+\mathrm{H}^{\vdot}\leftrightharpoons\mathrm{R}+\mathrm{H}_2$. Top: Geometry of the lTS resulting in a double-bonded dimer. Since the position (along the reaction path) of the lTS changes over temperature (highlighted in yellow in the inset), only the geometry at T=1200 K is shown. Bottom: Energy of the images along the MEP. More information about the temperature dependence of the position of the TS can be found in the Supplementary Information.}
    \label{fig:Reaction_3}
    \end{minipage}
    \hfill
    \begin{minipage}[t]{\columnwidth}
    \begin{center}
        \includegraphics[scale=0.22]{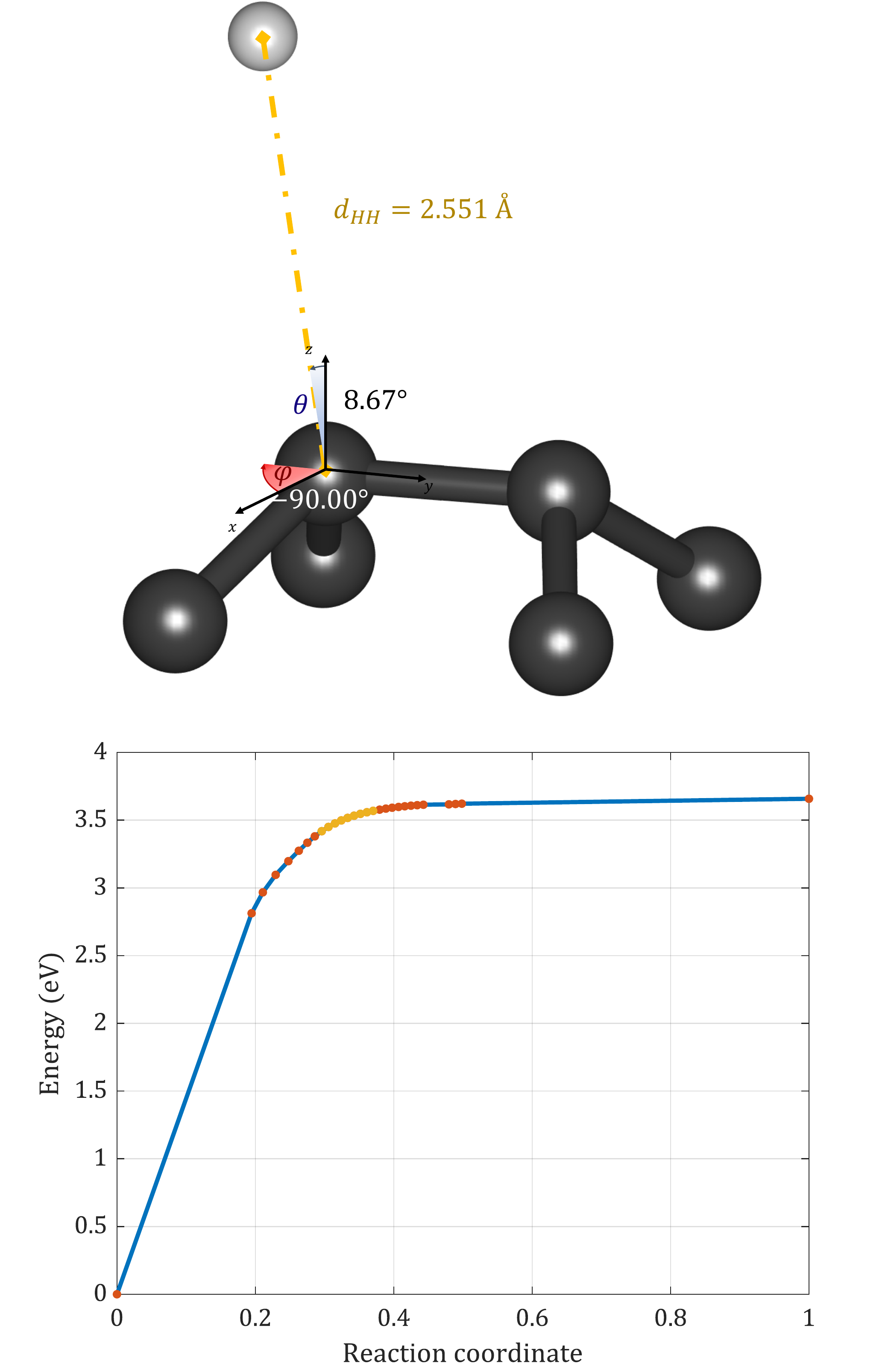}
    \end{center}
    \caption{Unimolecular dissociation of hydrogen on a half-passivated dimer (reaction 3 in Table \ref{tab:reaction_1-6}): $\mathrm{R^{\vdot}H}\leftrightharpoons\mathrm{R}+\mathrm{H}$. Top: Geometry of the lTS resulting in a double-bonded dimer. The position of the lTS changes over temperature: only the geometry corresponding to the lTS at T=1200 K is shown. Bottom: Energy of the images of the system during unimolecular dissociation of the remaining hydrogen on the dimer. Yellow dots highlight the positions of the lTS among the sampling point (red dots) from T=500 K to T=2500 K.}
    \label{fig:uni_diss}
    \end{minipage}
\end{figure*}

The unimolecular dissociation of an hydrogen atom on a fully-passivated dimer ($\mathrm{RH}_2\leftrightharpoons\mathrm{R^{\vdot}H}+\mathrm{H}^{\vdot}$, reaction 1) gives rise to a lTS, leading to effective reaction rate coefficients of $4.992\times10^{-3}$ s$^{-1}$ for the dissociation and $1.468\times10^{7}$ s$^{-1}$ for the recombination. The subsequent unimolecular dissociation of the remaining hydrogen atom on a half-passivated dimer ($\mathrm{R^{\vdot}H}\leftrightharpoons\mathrm{R}+\mathrm{H}^{\vdot}$, reaction 3) also gives rise to a lTS, which corresponding effective reaction rate coefficient are $1.790\times10^{1}$ s$^{-1}$ and $6.468\times10^{6}$ s$^{-1}$ for the dissociation and recombination respectively. The geometry of the lTS at T=1200 K along with the energy profile along the reaction path is shown in Fig. \ref{fig:uni_diss}.

As expected, for each reaction, we find values that are orders of magnitude different for the forward and backward directions, \textit{i.e.} the dissociation and recombination processes, respectively. All data can be found in Table \ref{tab:reaction_1-6}.

\subsection{\label{subsec:H_surf_cov}Hydrogen surface coverage}
Based on the reaction rate coefficients reported in table \ref{tab:reaction_1-6}, it is possible to compute the surface coverage fraction of the different types of dimers. In addition to the values of the prefactor $A_n$ and the energy of each TS, we also provide in table \ref{tab:reaction_1-6} the effective reaction rate coefficient $r_n$, that encompasses the influence of the gas concentration on the prevalence of each reaction and allows for comparison between gas-surface and surface-surface mechanisms at specific conditions of pressure, temperature and volume. Gas concentrations are calculated using ideal gas law, assuming a pressure of P=25 kPa, a temperature of $T=1200$ K, and a volume of the chamber of V=0.33 m$^3$, which are typical settings of experimental CVD diamond growth \cite{Rouzbahani2021}.

\begin{table*}[!htbp]
    \centering
        \caption{Summary of the TST parameters of the hydrogen ad/desorption processes at $T=1200$ K, P=25 kPa, V=0.33 m$^3$. $E_n^{TS}$ and $\Delta E_n$ are expressed in eV, $A_n$ and $k_n$ in units of s$^{-1}$ (\dag) or cm$^{3}\cdot$mol$^{-1}\cdot$s$^{-1}$ (\ddag) depending on the absence or presence of a gas reactant respectively, and $r_n$ in s$^{-1}$. Numbers in square brackets denote power of 10. All quantities are shown for the forward (Fwd) and backward (Bwd) directions of each reaction.}
        \label{tab:reaction_1-6}
        \begin{tabularx}{\textwidth}{c @{\extracolsep{\fill}}r@{\extracolsep{\fill}}c@{\extracolsep{\fill}}l@{\extracolsep{\fill}}c@{\extracolsep{\fill}}c@{\extracolsep{\fill}}c@{\extracolsep{\fill}}c@{\extracolsep{\fill}}c@{\extracolsep{\fill}}c@{\extracolsep{\fill}}c@{\extracolsep{\fill}}c}
            \hline
            \multicolumn{12}{|c|}{Hydrogen ad/desorption reactions}\\
            \hline
            \hline
            \hline
            \multirow{2}{*}{n} & \multicolumn{3}{c}{\multirow{2}{*}{Reaction}} & \multirow{2}{*}{TS} & \multirow{2}{*}{Direction} & \multirow{2}{*}{$\Delta E_n$ (eV)} & \multirow{2}{*}{$E_n^{TS}$ (eV)} & \multirow{2}{*}{$A_n$} & \multirow{2}{*}{$k_n$} & \multirow{2}{*}{units} &\multirow{2}{*}{$r_n$}\\ \\
            \hline
            \multirow{2}{*}{1}  & \multirow{2}{*}{$\mathrm{RH}_2$}
            & \multirow{2}{*}{$\leftrightharpoons$} & \multirow{2}{*}{$\mathrm{R^{\vdot}H}+\mathrm{H}^{\vdot}$} & \multirow{2}{*}{Loose}
                    & Fwd &  4.584 &  4.465 & 2.810[+16] & 4.992[$-$03] & \dag  & 4.992[$-$03]\\
            & & & & & Bwd & -4.584 & -0.120 & 1.844[+14] & 5.860[$+$14] & \ddag & 1.468[$+$07]\\
            \multirow{2}{*}{2}  & \multirow{2}{*}{$\mathrm{RH}_2+\mathrm{H}^{\vdot}$}
            & \multirow{2}{*}{$\leftrightharpoons$} & \multirow{2}{*}{$\mathrm{R^{\vdot}H}+\mathrm{H}_2$} & \multirow{2}{*}{Tight}
                    & Fwd &  0.052 &  0.091 & 5.908[+13] & 2.452[$+$13] & \ddag & 6.143[$+$05]\\
            & & & & & Bwd & -0.052 &  0.039 & 2.465[+11] & 1.697[$+$11] & \ddag & 4.252[$+$05]\\
            \multirow{2}{*}{3}  & \multirow{2}{*}{$\mathrm{R^{\vdot}H}$}
            & \multirow{2}{*}{$\leftrightharpoons$} & \multirow{2}{*}{$\mathrm{R}+\mathrm{H}^{\vdot}$} & \multirow{2}{*}{Loose}
                    & Fwd &  3.658 &  3.476 & 7.108[+15] & 1.790[$+$01] & \dag  & 1.790[$+$01]\\
            & & & & & Bwd & -3.658 & -0.182 & 4.442[+13] & 2.581[$+$14] & \ddag & 6.468[$+$06]\\
            \multirow{2}{*}{4}  & \multirow{2}{*}{$\mathrm{R^{\vdot}H}+\mathrm{H}^{\vdot}$}
            & \multirow{2}{*}{$\leftrightharpoons$} & \multirow{2}{*}{$\mathrm{R}+\mathrm{H}_2$} & \multirow{2}{*}{Loose}
                    & Fwd & -0.876 & -0.034 & 4.491[+13] & 1.343[$+$14] & \ddag & 3.364[$+$06]\\
            & & & & & Bwd &  0.876 &  0.842 & 1.801[+11] & 1.132[$+$08] & \ddag & 2.836[$+$02]\\
            \hline
        \end{tabularx}
\end{table*}


The very low values of the reaction rate coefficients for unimolecular dissociations---$k_1^F$=$4.992\times10^{-3}$ s$^{-1}$ and $k_3^F$=$1.790\times10^{1}$ s$^{-1}$ respectively the rates for forward direction of reactions $n$=1 and 3 in Table \ref{tab:reaction_1-6}---do not influence the steady-state concentrations and are ignored in the following. Therefore, we consider the only mean for the creation of non-passivated dimer to be reaction 4 in Table \ref{tab:reaction_1-6}.

To  calculate the steady-state concentration of $\mathrm{RH}_2$, $\mathrm{R^{\vdot}H}$ and $\mathrm{R}$, one must express the variations of their concentration:
\begin{widetext}
\begin{equation}
    \begin{aligned}
        \dv{\left[\mathrm{RH}_2\right]}{t}=&k_1^B\left[\mathrm{R^{\vdot}H}\right]\left[\mathrm{H}^{\vdot}\right]+k_2^B\left[\mathrm{R^{\vdot}H}\right]\left[\mathrm{H}_2\right]-k_2^F\left[\mathrm{RH}_2\right]\left[\mathrm{H}^{\vdot}\right],\\
        \dv{\left[\mathrm{R^{\vdot}H}\right]}{t}=&k_2^F\left[\mathrm{RH}_2\right]\left[\mathrm{H}^{\vdot}\right]+k_3^B\left[\mathrm{R}\right]\left[\mathrm{H}^{\vdot}\right]+k_4^B\left[\mathrm{R}\right]\left[\mathrm{H}_2\right]\\&-k_1^B\left[\mathrm{R^{\vdot}H}\right]\left[\mathrm{H}^{\vdot}\right]-k_2^B\left[\mathrm{R^{\vdot}H}\right]\left[\mathrm{H}_2\right]-k_4^F\left[\mathrm{R^{\vdot}H}\right]\left[\mathrm{H}^{\vdot}\right],\\
        \dv{\left[\mathrm{R}\right]}{t}=&k_4^F\left[\mathrm{R^{\vdot}H}\right]\left[\mathrm{H}^{\vdot}\right]-k_3^B\left[\mathrm{R}\right]\left[\mathrm{H}^{\vdot}\right]-k_4^B\left[\mathrm{R}\right]\left[\mathrm{H}_2\right].\\
    \end{aligned}
    \label{eq:variation_prim}
\end{equation}
\end{widetext}

Note that there are only two independent equations in this system, the second equation being a linear combination of the other two. At steady-state, these derivatives are zero. Furthermore, the $\mathrm{H}^{\vdot}$ concentration can be expressed as a fraction $f$ of the $\mathrm{H}_2$ concentration, which allows us to simplify the $\left[\mathrm{H}^{\vdot}\right]$ quantity that appear in all terms (experimentally, $\left[\mathrm{H}_2\right]$$=f\left[\mathrm{H}^{\vdot}\right]\simeq100\left[\mathrm{H}^{\vdot}\right]$ around T=1200 K \cite{Harris1990c}). If we also substitute $\left[\mathrm{RH}_2\right]$ with $1-\left[\mathrm{R^{\vdot}H}\right]-\left[\mathrm{R}\right]$, we can write the two remaining independent equations as:

\begin{equation}
    \begin{aligned}
        0&=k_2^F-\left[\mathrm{R^{\vdot}H}\right]\left(k_1^B+fk_2^B+k_2^F\right)-k_2^F\left[\mathrm{R}\right],\\
        0&=k_4^F\left[\mathrm{R^{\vdot}H}\right]-\left[\mathrm{R}\right]\left(k_3^B+fk_4^B\right).
    \end{aligned}
    \label{eq:syst_linear}
\end{equation}

At T=1200 K, such a linear system yields a steady-state surface coverage of 93.94\% of passivated dimers ($\mathrm{RH}_2$), 3.82\% of half-passivated dimers ($\mathrm{R^{\vdot}H}$) and 2.24\% of clean dimers ($\mathrm{R}$). These values agree well with the common experimental conception that the surface is almost entirely passivated, with a small fraction of $\mathrm{H}$ vacancies allowing for $\mathrm{CH}_3$ adsorption. For comparison, using gas phase reaction rate coefficients that were applied to model a diamond surface, the early work of Frenklach \cite{Frenklach1992} reported a value between 19 and 6.4\% for the half-passivated dimers, and between 11 and 1.5\% for the non-passivated dimers. A more recent work from May \textit{et al.} \cite{May2010} provides a value of about 10\%, but accounting for the presence of CH$_3$ while doing so, which can partially explain the discrepancy with our results.

\subsection{\label{subsec:Surface_radical(T,f)}Dependence of $\mathrm{H^{\vdot}}$ over gas radical mole fraction and temperature}
\begin{figure}[!h]
    \centering
    \includegraphics[scale=0.28]{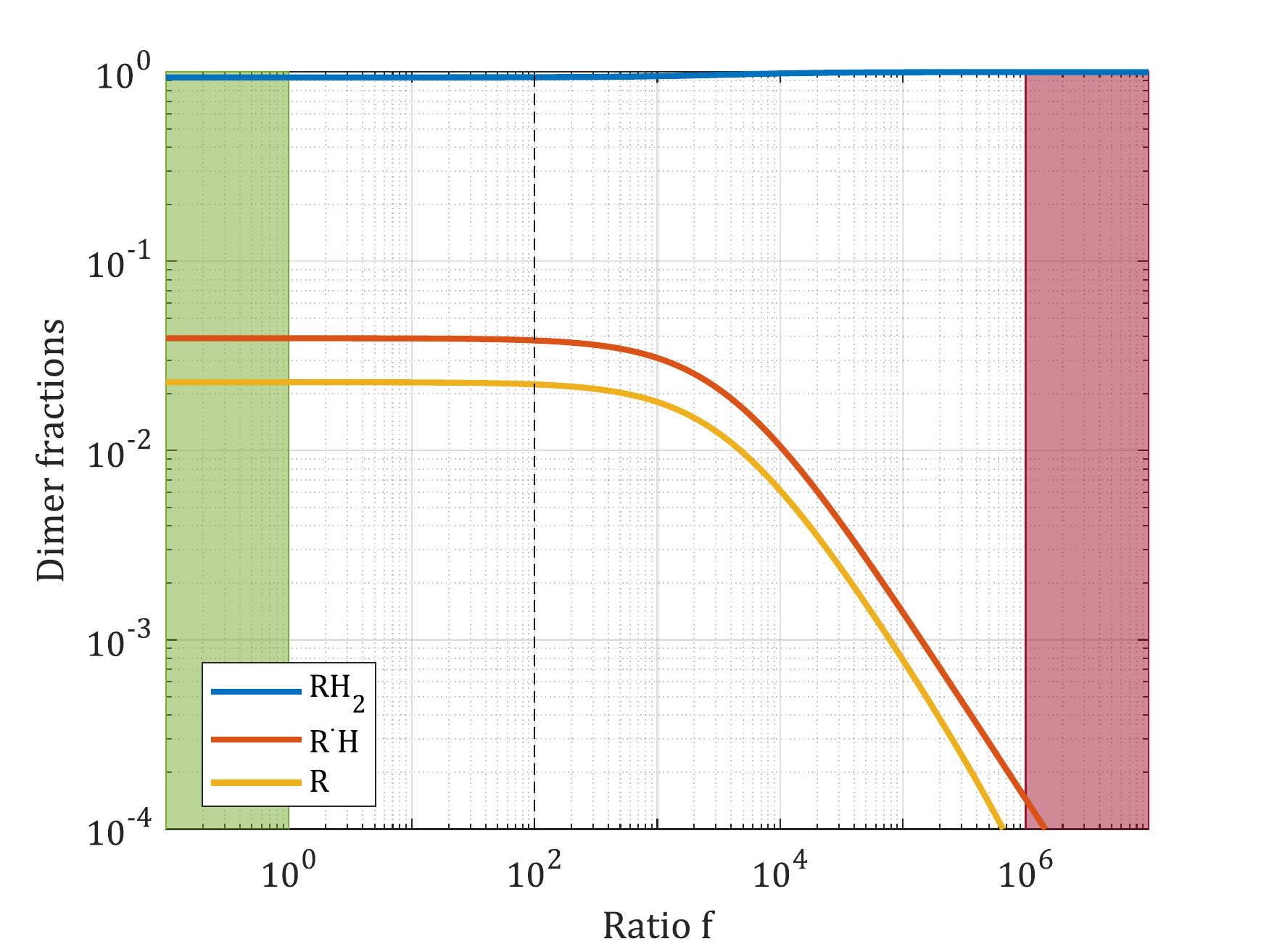}
    \caption{Evolution of the $\mathrm{RH}_2$, $\mathrm{R^{\vdot}H}$ and $\mathrm{R}$ concentrations with varying $f=\left[\mathrm{H}_2\right]/\left[\mathrm{H}^{\vdot}\right]$ ratio. Green area shows the asymptotic behaviour when there is a strong abundance of $\mathrm{H}^{\vdot}$, whereas the red area highlight the behaviour when $\mathrm{H}^{\vdot}$ are scarce.}
    \label{fig:f_var}
\end{figure}

This section aims at providing an insight into the dependence of the hydrogen surface coverage on the ratio $f=\left[\mathrm{H}_2\right]/\left[\mathrm{H}^{\vdot}\right]$. Varying this ratio between $f=10^{-1}$ and $f=10^{7}$ reveals two asymptotic trends highlighted by the green and red areas in Fig. \ref{fig:f_var}.

From an experimental point of view, closing the supply of H radicals---resulting in a H$_2$-only atmosphere---would drastically increase the $f$ ratio, prevent further H-desorption, thus ensuring passivation of the whole surface. As expected, our model driven by the system of linear equations (\ref{eq:syst_linear}) leads to a surface with fully passivated dimers once the H-radicals are no longer present in the gas phase. 

\begin{figure}[!h]
    \centering
    \includegraphics[scale=0.28]{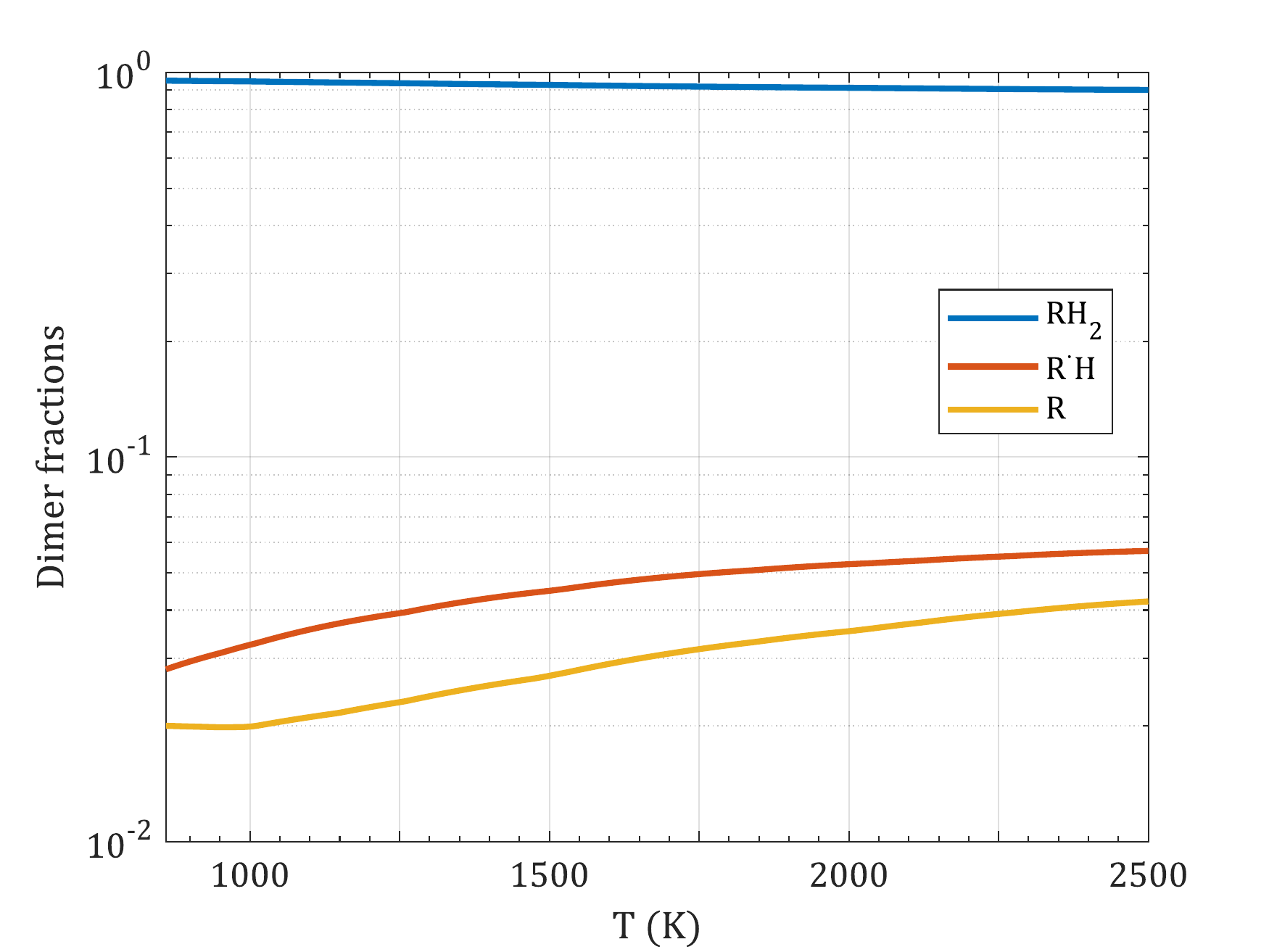}
    \caption{Evolution of the $\mathrm{RH}_2$, $\mathrm{R^{\vdot}H}$ and $\mathrm{R}$ concentrations between 860 K and 2500 K. As an approximation, the $f$-ratio is considered equal to 100 and constant over the whole range of temperature.}
    \label{fig:T_var}
\end{figure}

Assuming a constant concentration ratio $f$, an increase in temperature only lead to a slight increase of the half-passivated and clean dimers as depicted in Fig. \ref{fig:T_var}.

As $f$ is strongly related to the temperature of the chamber during CVD growth, an explicit temperature dependence of $f$ could extend this analysis to a broader range of temperatures. A description of this temperature dependent ratio $f$ could be achieved using a reliable gas-phase kinetic model, which is outside the scope of the current work. Detailed models of the gas phase inside a CVD reactor \cite{Truscott2016}, can pave the way for more refined calculations of the surface coverage of the surface.

\subsection{\label{subsec:H_mig}Migration of a hydrogen atom on a clean (100) surface}
Regardless of the theoretical or experimental approach, it is common to assume that the diamond surface is passivated by hydrogen. However, some extreme conditions (very low pressure and/or high temperature) may lead to the uncommon complete removal of hydrogen (these conditions cannot be modelled by the set of equations (\ref{eq:syst_linear}) as the assumption that unimolecular dissociation are negligible compared to gas-related interaction does not hold true at ultra-high vacuum pressures). 

\begin{figure}[!ht]
    \begin{center}
        \includegraphics[scale=0.58]{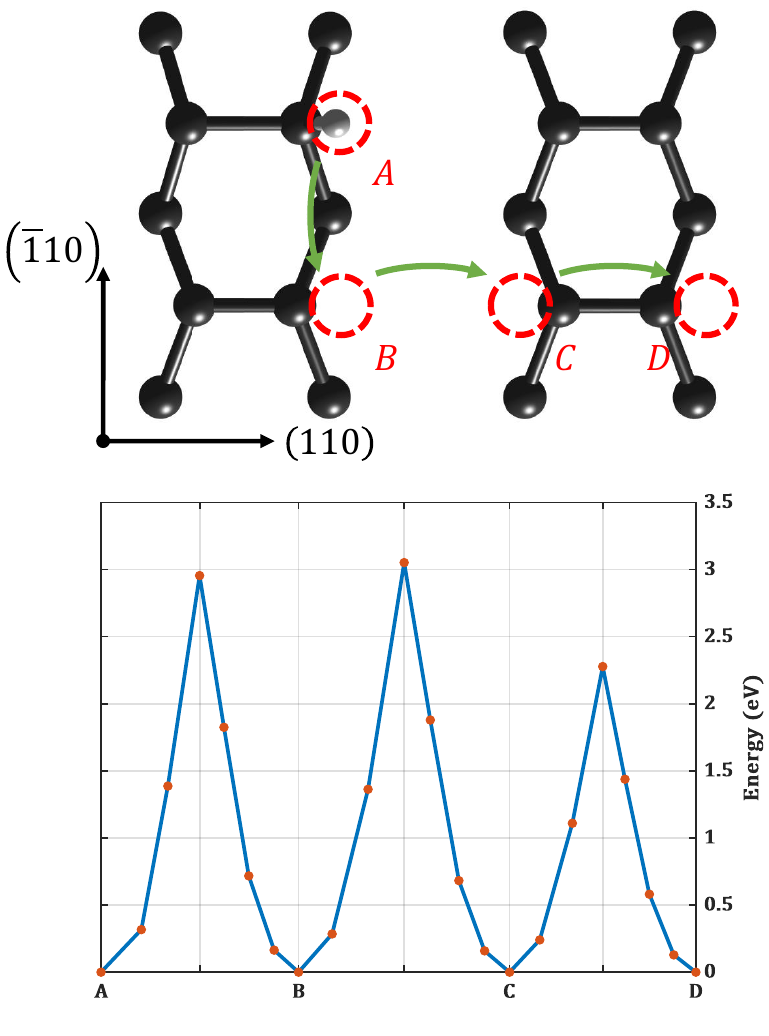}
        \caption{Mobility of a hydrogen adatom on a reconstructed, clean (100) diamond surface (reactions 5--7 in Table \ref{tab:E_Mob_H}). Top: Hydrogen atom moves from $A$ to $D$ via sites $B$ and $C$. Green arrows indicate the movement of the hydrogen atom, and red dashed circle point the location of sites A, B, C and D. Bottom: Energy of all the images along the migration of the H-adatom from $A$ to $D$.}
        \label{fig:Mob_H}
    \end{center}
\end{figure}

Therefore, we now investigate the migration of a lone hydrogen atom on a diamond (100) surface. The clean, non-H-passivated, (100) diamond surface shows a 2$\times$1 reconstruction similar to the passivated one. However, the dimers on the surface possess a different electronic configuration, which we earlier determined to be almost exclusively double bonded dimers ($\mathrm{R}$).

The detailed energy barriers for the migration of $\mathrm{H}$ on the surface (Fig. \ref{fig:Mob_H}) are reported in Table \ref{tab:E_Mob_H}. Migration from $A$ to $B$ (reaction 5) is a migration along dimer row, $B$ to $C$ (reaction 6) is a migration from one dimer row to another, and $C$ to $D$ (reaction 7) is a migration from one side of a dimer to the other. All intermediate steps along this path ($A$, $B$, $C$ and $D$) have the same total energy as they are equivalent by symmetry.

The large energy barriers for migration (greater than 2 eV) should not come as a surprise, as the distance between sites is about 2.5 \AA. Unsurprisingly, using TST, we report significantly low reaction rate coefficients for these migrations compared to gas-surface interactions discussed in Table \ref{tab:reaction_1-6}. 

\subsection{\label{subsec:H_vac_mig}Migration of hydrogen vacancies on a H-passivated (100) surface}
In previous studies \cite{Skokov1994b,Huang1992}, it was suggested that hydrogen vacancies (H-vacancies), both single and double, might migrate along dimer rows and columns on a passivated surface. To validate this assumption, the TST (see section \ref{subsec:TST}) is applied to determine the likelihood of each migration. Results are shown in Table \ref{tab:E_Mob_H}. In the single vacancy migration study (\textit{cf.} Fig. \ref{fig:Mob_H_vac}), the H-vacancy is moved along dimer row ($A$ to $B$, reaction 8), between two rows ($B$ to $C$, reaction 9) and between two sides of the same dimer ($C$ to $D$, reaction 10). 
\begin{table*}[!hbtp]
    \centering
        \caption{Summary of the TST parameters of the migrations of a hydrogen adatom on a clean surface, and that of a single or double H-vacancy on a passivated surface. Values have been computed at $T=1200$ K. $E_n^{TS}$ and $\Delta E_n$ are expressed in eV, and $A_n$ and $k_n$ in s$^{-1}$. Numbers in square brackets denote power of 10 and curly brackets indicate the range of energy barriers for different sets of calculation parameters from ref. \cite{Frenklach1997}. Reactions 5--10 are symmetrical (Sym.), whereas reactions 11--15 are asymmetrical: accordingly, $\Delta E_n$, $E_n^{TS}$, $A_n$ and $k_n$ are provided for the forward (Fwd) and backward (Bwd) directions.}
        \label{tab:E_Mob_H}
        \begin{tabularx}{\textwidth}{c@{\extracolsep{\fill}}c@{\extracolsep{\fill}}c@{\extracolsep{\fill}}c@{\extracolsep{\fill}} c@{\extracolsep{\fill}} c@{\extracolsep{\fill}} c@{\extracolsep{\fill}} c}
            \hline
            \multicolumn{8}{|c|}{\multirow{2}{*}{Migration of a hydrogen atom on a clean (100) surface (Fig. \ref{fig:Mob_H})}}\\
            \multicolumn{8}{|c|}{}\\
            \hline
            \multirow{2}{*}{n} & \multirow{2}{*}{Reaction} & & \multirow{2}{*}{Direction} & \multirow{2}{*}{$\Delta E_n$} & \multirow{2}{*}{$A_n$} & \multirow{2}{*}{$E_n^{TS}$} & \multirow{2}{*}{$k_n$}\\ \\
            \hline\\[-2ex]
            \multirow{1}{*}{5} & \multirow{1}{*}{$\mathrm{R}_{A}^{\vdot}\mathrm{H}+\mathrm{R}_{B}\leftrightharpoons\mathrm{R}_{A}+\mathrm{R}_{B}^{\vdot}\mathrm{H}$} & \multirow{1}{*}{This work} & \multirow{1}{*}{Sym.} & \multirow{1}{*}{0.000}
                  & \multirow{1}{*}{3.902[+14]} & \multirow{1}{*}{2.955} & \multirow{1}{*}{1.517[+02]}\\[1ex]
            \multirow{1}{*}{6} & \multirow{1}{*}{$\mathrm{R}_{B}^{\vdot}\mathrm{H}+\mathrm{R}_{C}\leftrightharpoons\mathrm{R}_{B}+\mathrm{R}_{C}^{\vdot}\mathrm{H}$} & \multirow{1}{*}{This work} & \multirow{1}{*}{Sym.} & \multirow{1}{*}{0.000}
                  & \multirow{1}{*}{6.123[+14]} & \multirow{1}{*}{3.052} & \multirow{1}{*}{9.334[+01]}\\[1ex]
            \multirow{1}{*}{7} & \multirow{1}{*}{$\mathrm{R}_{C}^{\vdot}\mathrm{H}\leftrightharpoons\mathrm{R}_{D}^{\vdot}\mathrm{H}$} & \multirow{1}{*}{This work} & \multirow{1}{*}{Sym.} & \multirow{1}{*}{0.000}
                  & \multirow{1}{*}{1.105[+14]} & \multirow{1}{*}{2.277} & \multirow{1}{*}{3.013[+04]}\\[1ex]
            \hline
            \multicolumn{8}{|c|}{\multirow{2}{*}{Migration of a single H-vacancy on a H-passivated (100) surface (Fig. \ref{fig:Mob_H_vac})}}\\
            \multicolumn{8}{|c|}{}\\
            \hline
            \multirow{2}{*}{n} & \multirow{2}{*}{Reaction} & & \multirow{2}{*}{Direction} & \multirow{2}{*}{$\Delta E_n$} & \multirow{2}{*}{$A_n$} & \multirow{2}{*}{$E_n^{TS}$} & \multirow{2}{*}{$k_n$}\\
            \\
            \hline\\[-2ex]
            \multirow{2}{*}{8} & \multirow{2}{*}{$\mathrm{R}_{A}^{\vdot}\mathrm{H}+\mathrm{R}_{B}\mathrm{H}_2\leftrightharpoons\mathrm{R}_{A}\mathrm{H}_2+\mathrm{R}_{B}^{\vdot}\mathrm{H}$} 
            & \multirow{1}{*}{This work}                 & \multirow{2}{*}{Sym.} &  \multirow{1}{*}{0.000} & \multirow{1}{*}{4.854[+14]} & \multirow{1}{*}{3.018} & \multirow{1}{*}{1.023[+02]}\\
            & & \multirow{1}{*}{Ref. \cite{Frenklach1997}} & &  \multirow{1}{*}{0.00}  & \multirow{1}{*}{--} & \multirow{1}{*}{\{2.90;3.82\}} & \multirow{1}{*}{--}\\
            \multirow{2}{*}{9} & \multirow{2}{*}{$\mathrm{R}_{B}^{\vdot}\mathrm{H}+\mathrm{R}_{C}\mathrm{H}_2\leftrightharpoons\mathrm{R}_{B}\mathrm{H}_2+\mathrm{R}_{C}^{\vdot}\mathrm{H}$} 
            & \multirow{1}{*}{This work}                 & \multirow{2}{*}{Sym.} &  \multirow{1}{*}{0.000} & \multirow{1}{*}{7.904[+14]} & \multirow{1}{*}{2.648} & \multirow{1}{*}{5.999[+03]}\\
            & & \multirow{1}{*}{Ref. \cite{Frenklach1997}} & &  \multirow{1}{*}{0.00}  & \multirow{1}{*}{--} & \multirow{1}{*}{2.88} & \multirow{1}{*}{--}\\
            \multirow{2}{*}{10} & \multirow{2}{*}{$\mathrm{R}_{C}^{\vdot}\mathrm{H}\leftrightharpoons\mathrm{R}_{D}^{\vdot}\mathrm{H}$} 
            & \multirow{1}{*}{This work}                 & \multirow{2}{*}{Sym.} &  \multirow{1}{*}{0.000} & \multirow{1}{*}{2.070[+14]} & \multirow{1}{*}{2.233} & \multirow{1}{*}{8.686[+04]}\\
            & & \multirow{1}{*}{Ref. \cite{Frenklach1997}} & &  \multirow{1}{*}{0.00}  & \multirow{1}{*}{--} & \multirow{1}{*}{\{2.11;3.06\}} & \multirow{1}{*}{--}\\[1ex]
            \hline
            \multicolumn{8}{|c|}{\multirow{2}{*}{Migration of a double H-vacancy on a H-passivated (100) surface (Fig. \ref{fig:Mob_H_vac2})}}\\
            \multicolumn{8}{|c|}{}\\
            \hline
            \multirow{2}{*}{n} & \multirow{2}{*}{Reaction} & & \multirow{2}{*}{Direction} & \multirow{2}{*}{$\Delta E_n$} & \multirow{2}{*}{$A_n$} & \multirow{2}{*}{$E_n^{TS}$} & \multirow{2}{*}{$k_n$}\\
            \\
            \hline\\[-2ex]
            \multirow{2}{*}{11} & \multirow{2}{*}{$\mathrm{R}_{\alpha}^{\vdot}\mathrm{H}+\mathrm{R}_{A}^{\vdot}\mathrm{H}+\mathrm{R}_{B}\mathrm{H}_2 \leftrightharpoons\mathrm{R}_{\alpha}^{\vdot}\mathrm{H}+\mathrm{R}_{A}\mathrm{H}_2+\mathrm{R}_{B}^{\vdot}\mathrm{H}$} & \multirow{2}{*}{This work} 
                  & \multirow{1}{*}{Fwd} & \multirow{1}{*}{-0.025} & \multirow{1}{*}{4.903[+14]} & \multirow{1}{*}{3.151} & \multirow{1}{*}{2.851[+01]}\\
              & & & \multirow{1}{*}{Bwd} &  \multirow{1}{*}{0.025} & \multirow{1}{*}{4.790[+14]} & \multirow{1}{*}{3.177} & \multirow{1}{*}{2.180[+01]}\\
            \multirow{2}{*}{12} & \multirow{2}{*}{$\mathrm{R}_{\alpha}^{\vdot}\mathrm{H}+\mathrm{R}_{B}^{\vdot}\mathrm{H}+\mathrm{R}_{C}\mathrm{H}_2 \leftrightharpoons\mathrm{R}_{\alpha}^{\vdot}\mathrm{H}+\mathrm{R}_{B}\mathrm{H}_2+\mathrm{R}_{C}^{\vdot}\mathrm{H}$}& \multirow{2}{*}{This work} 
                  & \multirow{1}{*}{Fwd} &  \multirow{1}{*}{0.012} & \multirow{1}{*}{5.600[+14]} & \multirow{1}{*}{2.719} & \multirow{1}{*}{2.130[+03]}\\
              & & & \multirow{1}{*}{Bwd} & \multirow{1}{*}{-0.012} & \multirow{1}{*}{5.701[+14]} & \multirow{1}{*}{2.707} & \multirow{1}{*}{2.428[+03]}\\
            \multirow{2}{*}{13} & \multirow{2}{*}{$\mathrm{R}_{\alpha}^{\vdot}\mathrm{H}+\mathrm{R}_{C}^{\vdot}\mathrm{H} \leftrightharpoons\mathrm{R}_{\alpha}^{\vdot}\mathrm{H}+\mathrm{R}_{D}^{\vdot}\mathrm{H}$}& \multirow{2}{*}{This work} 
                  & \multirow{1}{*}{Fwd} & \multirow{1}{*}{-0.017} & \multirow{1}{*}{1.051[+15]} & \multirow{1}{*}{2.563} & \multirow{1}{*}{1.813[+03]}\\
              & & & \multirow{1}{*}{Bwd} &  \multirow{1}{*}{0.017} & \multirow{1}{*}{1.028[+14]} & \multirow{1}{*}{2.580} & \multirow{1}{*}{1.512[+03]}\\
            \multirow{4}{*}{14} & \multirow{4}{*}{$\mathrm{R}^{\vdot}_{\alpha}\mathrm{H}+\mathrm{R}_{D}^{\vdot}\mathrm{H} \leftrightharpoons\mathrm{R}_{\alpha,E}+\mathrm{R}_D\mathrm{H}_2$}& \multirow{2}{*}{This work}
                  & \multirow{1}{*}{Fwd} & \multirow{1}{*}{-0.924} & \multirow{1}{*}{2.954[+14]} & \multirow{1}{*}{2.321} & \multirow{1}{*}{5.274[+04]}\\
              & & & \multirow{1}{*}{Bwd} &  \multirow{1}{*}{0.924} & \multirow{1}{*}{7.117[+14]} & \multirow{1}{*}{3.245} & \multirow{1}{*}{1.669[+01]}\\
            & & \multirow{2}{*}{Ref. \cite{Frenklach1997}}  
                  & \multirow{1}{*}{Fwd} & \multirow{1}{*}{\{-0.01;0.05\}} & \multirow{1}{*}{--} & \multirow{1}{*}{\{2.96;4.80\}} & \multirow{1}{*}{--} \\
              & & & \multirow{1}{*}{Bwd} & \multirow{1}{*}{--} & \multirow{1}{*}{--} & \multirow{1}{*}{--} & \multirow{1}{*}{--}\\
            \multirow{4}{*}{15} & \multirow{4}{*}{$\mathrm{R}_{\alpha,E}+\mathrm{R}_F\mathrm{H}_2\leftrightharpoons\mathrm{R}^{\vdot}_{\alpha}\mathrm{H}+\mathrm{R}^{\vdot}_{F}\mathrm{H}$}
            & \multirow{2}{*}{This work} 
                  & Fwd &  \multirow{1}{*}{0.923} & \multirow{1}{*}{1.015[+15]} & \multirow{1}{*}{3.437} & \multirow{1}{*}{3.725[+00]} \\
              & & & Bwd & \multirow{1}{*}{-0.923} & \multirow{1}{*}{4.211[+14]} & \multirow{1}{*}{2.514} & \multirow{1}{*}{1.161[+04]} \\
            & & \multirow{2}{*}{Ref. \cite{Frenklach1997}}
                  & Fwd & \multirow{1}{*}{0.05} & \multirow{1}{*}{--} & \multirow{1}{*}{2.89} & \multirow{1}{*}{--}\\
              & & & Bwd & \multirow{1}{*}{--} & \multirow{1}{*}{--} & \multirow{1}{*}{--} & \multirow{1}{*}{--}\\
            \hline
        \end{tabularx}
\end{table*}
\begin{figure*}[!htbp]
    \begin{center}
        \includegraphics[scale=0.56]{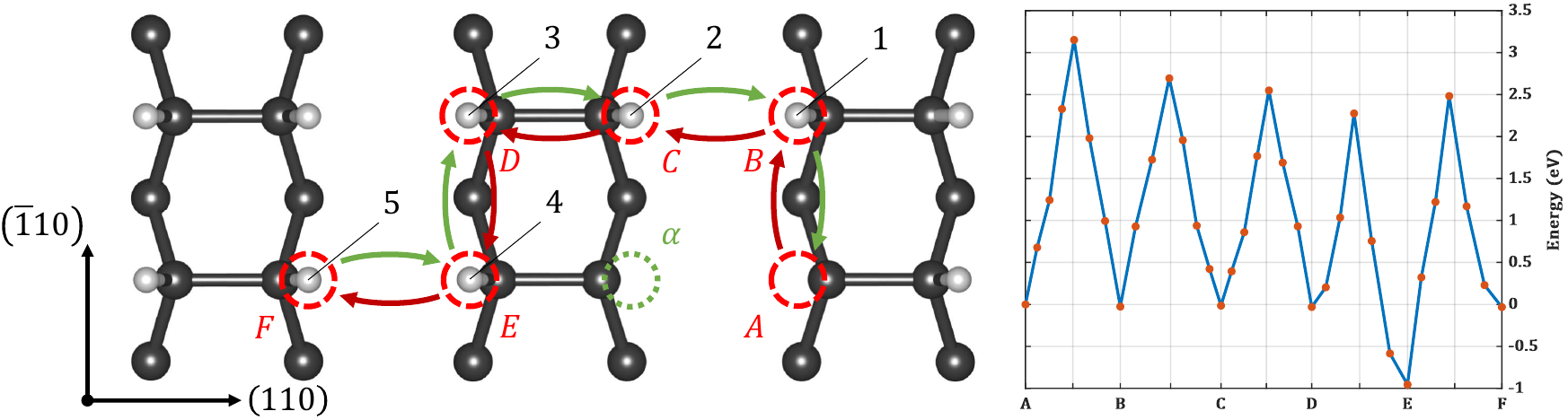}
        \caption{Mobility of a H-vacancy around a fixed H-vacancy $\alpha$ on a passivated (100) diamond surface  (reactions 11--15 in Table \ref{tab:E_Mob_H}). Left: Hydrogen atom $1$ moves from $B$ to $A$ thus moving the vacancy from $A$ to $B$, while the vacancy at $\alpha$ stands still (reaction 11). Vacancy is moved the same way from $B$ to $C$  (reaction 12), from $C$ to $D$ (reaction 13), from $D$ to $E$ (reaction 14), and from $E$ to $F$ (reaction 15). Green arrows: movement of the hydrogen atoms. Red arrows: movement of the vacancy. Right: Energy of the images along the migration of the H-vacancy from $A$ to $F$.}
        \label{fig:Mob_H_vac2}
    \end{center}
\end{figure*}

\begin{figure}[!h]
    \begin{center}
        \includegraphics[scale=0.56]{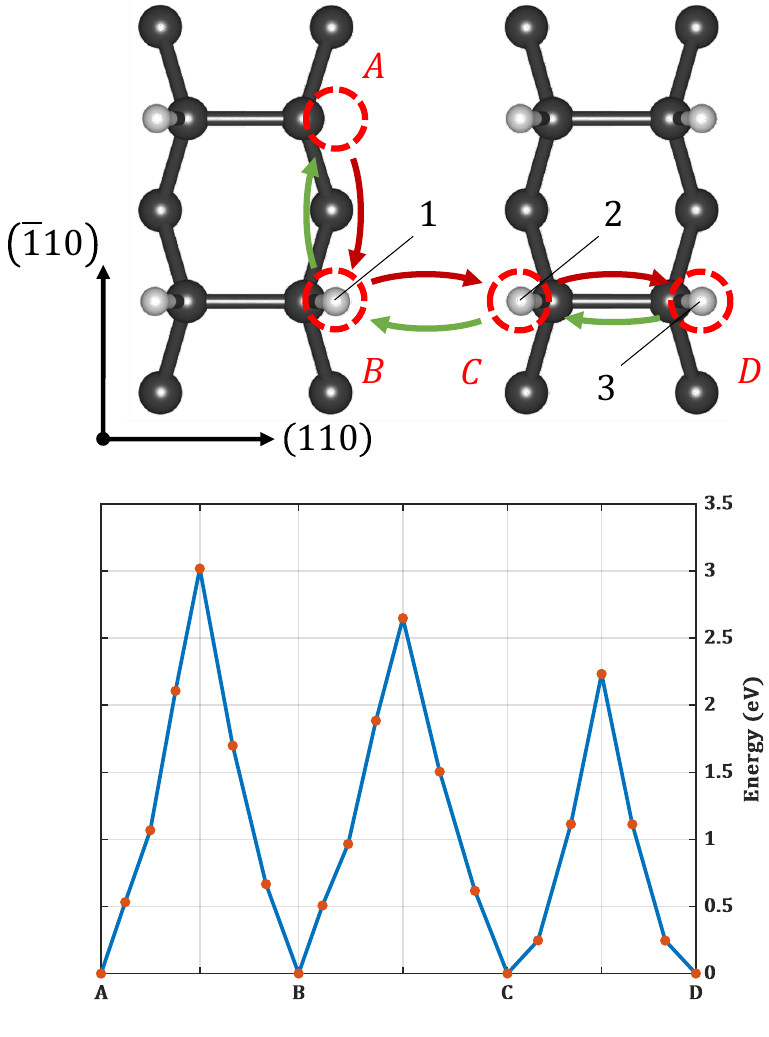}
        \caption{Mobility of a H-vacancy on a passivated (100) diamond surface (reactions 8--10 in Table \ref{tab:E_Mob_H}). Top: Hydrogen atom $1$ initially at $B$ moves to $A$ thus moving the vacancy from $A$ to $B$. The same process is employed to move the vacancy further from $B$ to $C$ and from $C$ to $D$. Green arrows indicate the movement of the hydrogen atoms and red arrows indicate the movement of the vacancy. Bottom: Energy of all images along the migration of the H-vacancy from $A$ to $D$.}
        \label{fig:Mob_H_vac}
    \end{center}
\end{figure}

Among these single-vacancy migration reactions, the movement of a vacancy between two sides of a dimer ($C$ to $D$, reaction 10, $r_{10}=8.686\cdot10^{4}$ s$^{-1}$) is the fastest one: its reaction rate coefficient is comparable with the ad/desorption mechanisms reported in Table \ref{tab:reaction_1-6}. However, it has little importance on the overall movement of vacancies on a passivated surface as it remains on a given dimer. Indeed, to move on the surface, the vacancy must migrate along ($r_{8}=1.023\times10^{2}$ s$^{-1}$) or between dimer rows ($r_{9}=5.999\times10^{3}$ s$^{-1}$). The latter reaction rate coefficient turns out to be about 58 times faster, which suggests a net surface migration anisotropy. Along with the even higher reaction rate coefficient of the movement between two sides of a dimer, this anisotropy favours vacancy migration across dimer rows rather than along a specific row. 

We also investigate the influence of a pre-existing vacancy defect on the migration of a second vacancy to assess whether migrations of vacancies towards each other are likely phenomena. A similar migration pattern as before is used. One H-vacancy is kept frozen at a given position ($\alpha$ in Fig. \ref{fig:Mob_H_vac2}) while the other vacancy moves along another dimer row ($A$ to $B$, reaction 11), then to the row hosting the fixed vacancy ($B$ to $C$, reaction 12). Once both vacancies are on the same dimer row, the moving vacancy travels towards the frozen vacancy via sites $D$ (reaction 13) and $E$ (reaction 14), after which it is moved away from $\alpha$ towards site $F$ (reaction 15). 
Just as previously, this results in large energy barriers for the vacancy migration, as shown on Fig. \ref{fig:Mob_H_vac2}.

Table \ref{tab:E_Mob_H} shows that a frozen vacancy does not influence much the TS energies of the migration of a distant H-vacancy (see the difference between reactions 8 and 11, 9 and 12, and 10 and 13). However, vacancies on adjacent $\mathrm{R^{\vdot}H}$ dimers are more likely to join and create a clean dimer than to drift away from each other, as demonstrated by the higher reaction rate coefficient coefficient $k_n$ for the forward reaction 14 ($r_{14}=5.274\times10^{4}$ s$^{-1}$) and the backward reaction 15 ($r_{15}=1.161\times10^{4}$ s$^{-1}$), \textit{i.e.} for the formation of a double-bonded dimer by means of migration.

The configuration resulting from reaction 14 exhibits a vacancy on each side of a dimer, that is on sites $\alpha$ and $E$, leaving a clean double-bonded dimer. As previously discussed, the energy of this spin configuration is about 1 eV lower than that of the single-bonded dimer configuration, due to the formation of a double-bond. The migration leading to this clean double-bonded dimer can be seen as an alternative reaction mechanism to the radical attack (Figs. \ref{fig:Reaction_3}) to produce double-bonded, non-passivated dimers. This alternative reaction mechanism can alter the fractions of the different dimers shown in Figs. \ref{fig:f_var} and \ref{fig:T_var}: the adjacent dimers accounting for about 3.82\%$\times$3.82\%=0.15\% of the surface dimers---and assuming that the other migration processes occurs on a much longer time-scale than the gas-surface processes---it would change the fraction of non-passivated dimer from 2.24\% to 2.39\%. Note that this is only a rough approximation of the change that could be brought to the surface fractions: depending on the temperature and pressure, these migrations have an even smaller minor impact on the conclusions resulting from our model since the effective reaction rate coefficient values of the migrations creating clean dimers are significantly lower than the reaction rate coefficients of the gas-surface interactions. A more detailed calculation of the influence of the migrations on the surface coverage fraction wouldrequire a different approach than the one discussed in \ref{subsec:H_surf_cov}.

\section{\label{sec:level4}Conclusion}
This work reflects on the important role played by hydrogen radicals to create reactive surface sites during diamond growth. In the context of CVD growth, these active sites determine the ability for methyl radicals to bond to the surface that eventually leads to the deposition of additional layer of diamond. Therefore, their concentration is a crucial parameter during diamond growth.

We present a comprehensive description of hydrogen radicals and molecules exchange between the surface sites and between the surface and the atmosphere. The importance of the inclusion of vdW corrections to the DFT calculations are emphasised in Fig. 2 where we show that all vdW correction schemes leads to significantly lower energy barriers than non-vdW calculations. For the loose TS of reaction 3, the distance between the hydrogen and the surface is in the range where one expects the influence of vdW corrections to be significant, which in turn affects the vibrational frequencies used to compute the reaction rate coefficients. The absence of previous ab-initio calculations of reaction rates (including the Arrhenius prefactor) prevent us from making any definitive conclusion on the influence vdW corrections have on the results presented in this work. The methods we use in this work---DFT, cNEB and VTST---allow to compare likelihood of different reactions: the first hydrogen desorption (reaction 2 in Table \ref{tab:reaction_1-6}) constitutes a compelling illustration of the respective importance of the different factors influencing the effective reaction rate coefficient $r_n$: (1) the exponential prefactor, (2) the energy barrier and (3) the concentration of the radical/molecule species involved in the reaction:

\begin{enumerate}
    \item As shown in section \ref{subsec:TST}, the exponential prefactor $A_n$ depends on the partition function of the TS but also on the partition function of the initial or final state (for the forward or backward reactions respectively). For this representative reaction, the exponential prefactors suggest the forward ($5.908\cdot10^{13}$ cm$^3\vdot$mol$^{-1}\vdot$s$^{-1}$) is more likely than the backward ($2.465\cdot10^{11}$ cm$^3\vdot$mol$^{-1}\vdot$s$^{-1}$). 
    \item The calculated energy barriers $\Delta E$ using the cNEB method differs for the forward (0.091 eV) and backward (0.039 eV) reactions, favoring the latter over the former.
    \item Under the assumption of a $\mathrm{H}^{\cdot}$ molar concentration of 1\% and a $\mathrm{H}_2$ molar concentration of 99\%,\cite{Harris1990c} the backward reaction involving $\mathrm{H}_2$ is favoured over the forward involving $\mathrm{H}^{\cdot}$.
\end{enumerate}

This emphasises the importance of a complete description of the reaction: taken separately, the three points highlighted above can lead to diverging conclusion. We report lower values for the energy of the tTS associated with the first hydrogen desorption and make a clear distinction between the latter reaction and the second desorption, \textit{i.e.} the removal of the remaining hydrogen on a half-passivated dimer. As a matter of fact, the two successive reactions that form the clean dimers turns out to be intrinsically different as one occurs through a tTS at the second through a lTS, each leading to a specific reaction rate coefficient, hence the need for a robust approach as the one presented herein that is able to compute reaction rate coefficients whether their energy profile shows a barrier or not. The importance of atomic-scale calculations lies in the detailed description of the geometry of the TS, of the energy profile of the reaction and of its reaction rate coefficients and reaction rate coefficients. To the best of our knowledge, no previous theoretical study computed reaction rates coefficients of hydrogen-related reactions on hydrogen passivated (100) diamond surfaces at the GGA level of the density functional theory, including spin-polarisation effect and considering long-range interactions corrections (vdW dispersion).

This comprehensive study is the key to build a stochastic or mechanical \cite{DiTeodoro2022} model that estimates the surface coverage of the different types of dimers (fully/half-passivated and clean dimers) as a function of temperature and gas radical concentrations. Our simple model provides the surface coverage \textit{during} the exposure to the $\mathrm{H}^{\vdot}$-rich atmosphere, which---to the best of our knowledge---cannot be measured experimentally.

Additionally, we quantify a net surface vacancy migration anisotropy, that favours migrations across dimer rows. Furthermore, we report that most migration reactions do not compete with gas-surface interactions at usual CVD diamond growth conditions. However, the latter does not hold true for adjacent half-passivated dimers: the only two reactions that almost compete with gas-surface interactions favours the production of the non-reactive, clean dimers over the reactive half-passivated dimers. To some extent, these reactions temper the surface coverage of the different dimers provided by our simple model, which in its present state does not account for this particular effect. The amplitude of this effect has yet to be exactly determined, but a first approximation indicates that it is small: non-passivated dimer fraction would increase from 2.24\% to 2.39\% at usual CVD conditions.

We are confident that the methods presented here are relevant for studying other mechanisms such as the nucleation on a (100) diamond surface as well as other crystallographic orientations or even other semi-conductors.

\section*{Prime novelty statement}
The radical attack of hydrogen on the (100) H-passivated surface is investigated by means of first principles calculations. A comprehensive set of reaction rate coefficients (including energy barriers and profiles of each reaction) is calculated based on quantum mechanical calculations of the energies and vibrational spectrum. The role of electron spin is elucidated, and H surface coverage of fully passivated, half-passivated and clean dimers is estimated. The importance of hydrogen migrations reactions is also assessed: only two mechanisms can compete with gas-surface interactions and temper the relative distribution of each species.
\section*{CRedit authorship contribution statement}
\textbf{Emerick Y. Guillaume:}       Methodology, Investigation, Software, Formal analysis, Visualization , Writing - Original Draft, Writing - Review \& Editing
\textbf{Danny E. P. Vanpoucke:}      Conceptualization, Writing - Review \& Editing, Supervision
\textbf{Rozita Rouzbahani:}          Writing - Review \& Editing
\textbf{Luna Pratali Maffei:}        Methodology, Writing - Review \& Editing
\textbf{Matteo Pelucchi:}            Methodology, Writing - Review \& Editing, Supervision
\textbf{Yoann Olivier:}              Writing - Review \& Editing, Supervision
\textbf{Luc Henrard:}                Conceptualization, Writing - Review \& Editing, Supervision
\textbf{Ken Haenen:}                 Conceptualization, Writing - Review \& Editing, Supervision

\section*{Declaration of competing interest}
The authors declare that they have no known competing financial interests or personal relationship that could have appeared to influence the work reported in this paper.

\section*{\label{sec:level5}Acknowledgment}
EYG's scholarship is funded by ADRE and BOF under grant no. A-8581 and R-9982. The resources and services used in this work were provided by the VSC (Flemish Supercomputer Center), funded by the Research Foundation - Flanders (FWO) and the Flemish Government, and by the Consortium des Équipements de Calcul Intensif (CÉCI), funded by the Fonds de la Recherche Scientifique de Belgique (F.R.S.-FNRS) under Grant No. 2.5020.11 and by the Walloon Region. LPM's contribution to this study was carried out within the NEST - Network 4 Energy Sustainable Transition (D.D. 1243 02/08/2022, PE00000021) and received funding under the National Recovery and Resilience Plan (NRRP), Mission 4 Component 2 Investment 1.3, funded from the European Union - NextGenerationEU. This manuscript reflects only the author's views and opinions, neither the European Union nor the European Commission can be considered responsible for them. Finally, this work was also financially supported by the Methusalem NANO network. EYG would like to thank P. Pobedinskas for the in-depth discussions on the experimental growth of diamond.
\nocite{molecule3D}
\bibliography{Biblio}

\end{document}


\preprint{APS/123-QED}
\title{Supplementary Information:\\First-principles investigation of hydrogen-related reactions\\on (100)--(2$\times$1)$:$H diamond surfaces}

\author{Emerick Y. Guillaume}
\email[Corresponding author: ]{Emerick.Guillaume@UHasselt.be}
\affiliation{Namur Institute of Structured Matter (NISM), University of Namur, Rue de Bruxelles 61, 5000 Namur, Belgium}
\affiliation{Hasselt University, Institute for Materials Research (IMO-IMOMEC), 3590 Diepenbeek, Belgium}
\affiliation{IMOMEC, IMEC vzw, Wetenschapspark 1, 3590 Diepenbeek, Belgium}
\author{Danny E. P. Vanpoucke}
\affiliation{Hasselt University, Institute for Materials Research (IMO-IMOMEC), 3590 Diepenbeek, Belgium}
\affiliation{IMOMEC, IMEC vzw, Wetenschapspark 1, 3590 Diepenbeek, Belgium}
\author{Rozita Rouzbahani}
\affiliation{Hasselt University, Institute for Materials Research (IMO-IMOMEC), 3590 Diepenbeek, Belgium}
\affiliation{IMOMEC, IMEC vzw, Wetenschapspark 1, 3590 Diepenbeek, Belgium}
\author{Luna Pratali Maffei}
\affiliation{CRECK Modeling Lab, Politecnico di Milano, Piazza L. da Vinci, 32, 20133 Milano, Italy}
\author{Matteo Pelucchi}
\affiliation{CRECK Modeling Lab, Politecnico di Milano, Piazza L. da Vinci, 32, 20133 Milano, Italy}
\author{Yoann Olivier}
\affiliation{Namur Institute of Structured Matter (NISM), University of Namur, Rue de Bruxelles 61, 5000 Namur, Belgium}
\author{Luc Henrard}
\affiliation{Namur Institute of Structured Matter (NISM), University of Namur, Rue de Bruxelles 61, 5000 Namur, Belgium}
\author{Ken Haenen}
\affiliation{Hasselt University, Institute for Materials Research (IMO-IMOMEC), 3590 Diepenbeek, Belgium}
\affiliation{IMOMEC, IMEC vzw, Wetenschapspark 1, 3590 Diepenbeek, Belgium}
\date{\today}

\maketitle

\onecolumngrid
\section{Surface energy convergence}
We model the diamond surface as a periodic 11-layers slab (containing about 208 atoms), which unit cell is defined by a 10.06$\times$10.06$\times$20.89 \AA $ $ orthorombic unit cell. Calculation were performed using the PBE implementation of the GGA in VASP, with an energy cut-off of 650 eV and a 4x4x1 k-point sampling of the Brillouin zone. These parameters were chosen by ensuring convergence of the surface energy of a primitive surface cell of 2.51$\times$2.51$\times$20.89 \AA. Since a 7x7x1 k-point sampling was found sufficient to converge this small surface cell, increasing this cell to a 10.06$\times$10.06$\times$20.89 \AA $ $ lowers the k-point sampling requirement. All convergence test are shown in figure \ref{fig:conv_test}. The energy barrier of reaction 2 was computed with two set of k-point, $\Gamma$-only and 4$\times$4$\times$1, and no significant difference were found as highlighted in Table I of the main paper. Due to the important computational cost of vibrational frequencies calculations, they were perform at the $\Gamma$-only level.

\section{Spin-dependent reaction paths}
As stated in the main manuscript, the cNEB method do not permit spin configuration change during the reaction. As such, for a more exhaustive understanding of the reactions leading to a non-passivated dimer, we also considered the triplet state configuration of the non-passivated dimer to be a possible product. Indeed, depending on the spin-configuration of the reactant, non-passivated dimers can be found in a single- or double-bonded form. For the sake of brevity, single-bonded dimers are referred to as $\mathrm{R}^{2\vdot}$ since they are bi-radical sites and double-bonded $\mathrm{R}$ as they are not radicals.

In contrast with singlet state production (reaction 4), the reaction leading to a single-bonded dimer (reaction 4') possesses a tTS. The TST methodology thus applies and results in the following effective reaction rate coefficients: $4.246\times10^{5}$ s$^{-1}$ and $5.383\times10^{5}$ s$^{-1}$ for the forward and backward reaction, respectively.

The fact that the reaction rate coefficient for the reaction leading to a double-bonded dimer is about 10 times larger than the one leading to the single-bonded dimer suggests that the latter can only play a minor role in the processes happening on the surface. Although the two reactions start with the same reactants, their spin configuration are different: double-bonded dimers arise from a singlet configuration, whereas single-bonded ones arise from a triplet configuration. If the energies of the two configurations are identical in the reactants, this fact does not hold true all along the reaction path and intersystem crossing (ISC) might play a role through conversion of the reactant from a triplet to a singlet state as this configuration gets increasingly favoured by its energy profile, as shown in Fig. 3 of the main manuscript and in Fig. \ref{fig:SI_reaction_4_prim} of this SI. We observe a similar issue with the unimolecular dissociation, which effective reaction rate coefficient are $2.538\times10^{-3}$ s$^{-1}$ and $1.490\times10^{7}$ s$^{-1}$ for the dissociation of $\mathrm{R^{\vdot}H}$ and the recombination of $\mathrm{R}$ with $\mathrm{H}$ respectively. A detailed description of the TST parameters can be found in Table \ref{tab:triplet_processes}.

As shown in section \ref{subsec:pi_over_sigma}, virtually all dimers will eventually be found in their singlet state. We highlighted previously that there is a ten-fold difference between the rate coefficient of reaction 4 and 4$'$ that lead to double- and single-bonded dimers, respectively. Reaction 4$'$ even becomes negligible depending on the ISC rate coefficient between singlet and triplet \textit{during} the reaction: if the ISC rate coefficient is faster than reaction 4 and 4$'$, then the only means of production of non-passivated dimer through radical attack is reaction 4. However, if the ISC rate coefficient is slower, then reaction 4$'$ also contributes to the production of non-passivated dimers via production of short-lived triplet dimers. Given the absence of a conclusive statement regarding the relative speed of the ISC and reaction 4 and 4$'$, we follow the broad consensus over the fact that ISC is a fast process, therefore eliminating the need to consider triplet state dimer production in the hydrogen surface coverage calculations.

\begin{table*}[!htbp]
    \centering
        \caption{TST parameters of the hydrogen ad/desorption reactions producing single-bonded dimers at $T=1200$ K, P=25 kPa, V=0.33 m$^3$. $E_n^{TS}$ and $\Delta E_n$ are expressed in eV, $A_n$ and $k_n$ in units of s$^{-1}$ (\dag) or cm$^{3}\cdot$mol$^{-1}\cdot$s$^{-1}$ (\ddag) depending on the absence or presence of a gas reactant respectively, and $r_n$ in s$^{-1}$. Numbers in square brackets denote power of 10. All quantities are shown for the forward (Fwd) and backward (Bwd) directions of each reaction.}
        \label{tab:reaction_1-6}
        \begin{tabularx}{\textwidth}{c @{\extracolsep{\fill}}r@{\extracolsep{\fill}}c@{\extracolsep{\fill}}l@{\extracolsep{\fill}}c@{\extracolsep{\fill}}c@{\extracolsep{\fill}}c@{\extracolsep{\fill}}c@{\extracolsep{\fill}}c@{\extracolsep{\fill}}c@{\extracolsep{\fill}}c@{\extracolsep{\fill}}c}
            \hline
            \multicolumn{12}{|c|}{Hydrogen ad/desorption reactions}\\
            \hline
            \hline
            \hline
            \multirow{2}{*}{n} & \multicolumn{3}{c}{\multirow{2}{*}{Reaction}} & \multirow{2}{*}{TS} & \multirow{2}{*}{Direction} & \multirow{2}{*}{$\Delta E_n$ (eV)} & \multirow{2}{*}{$E_n^{TS}$ (eV)} & \multirow{2}{*}{$A_n$} & \multirow{2}{*}{$k_n$} & \multirow{2}{*}{units} &\multirow{2}{*}{$r_n$}\\ \\
            \hline
            \multirow{2}{*}{3$'$}  & \multirow{2}{*}{$\mathrm{R^{\vdot}H}$}
            & \multirow{2}{*}{$\leftrightharpoons$} & \multirow{2}{*}{$\mathrm{R}^{2\vdot}+\mathrm{H}^{\vdot}$} & \multirow{2}{*}{Loose}
                    & Fwd &  4.695 &  4.571 & 4.016[+16] & 2.538[$-$03] & \dag  & 2.538[$-$03]\\
            & & & & & Bwd & -4.695 & -0.124 & 1.802[+14] & 5.947[$+$14] & \ddag & 1.490[$+$07]\\
            \multirow{2}{*}{4$'$}  & \multirow{2}{*}{$\mathrm{R^{\vdot}H}+\mathrm{H}^{\vdot}$}
            & \multirow{2}{*}{$\leftrightharpoons$} & \multirow{2}{*}{$\mathrm{R}^{2\vdot}+\mathrm{H}_2$} & \multirow{2}{*}{Tight}
                    & Fwd &  0.163 &  0.166 & 8.396[+13] & 1.695[$+$13] & \ddag & 4.246[$+$05]\\
            & & & & & Bwd & -0.163 & -0.003 & 2.201[+11] & 2.148[$+$11] & \ddag & 5.383[$+$05]\\
            \hline
        \end{tabularx}
        \label{tab:triplet_processes}
\end{table*}

\section{\label{subsec:pi_over_sigma}Relative distribution of single- and double-bonded dimers}
On a non-passivated diamond surface in vacuum, we assume there exist steady-state concentrations of single and double bond dimers ($\left[\mathrm{R}^{2\vdot}\right]$ and $\left[\mathrm{R}\right]$). The only reactions that can occur under such circumstances are the transformation of one dimer type to the other, characterised by the reaction rate coefficient $k_{d\rightarrow s}^F$ and $k_{d\rightarrow s}^B$ for the transformation of double ($d$) to single ($s$) bond dimers and conversely. Variations of concentrations are written as:
\begin{equation}
    \begin{aligned}
        \dv{t}\left[\mathrm{R}\right]=&k_{d\rightarrow s}^B\left[\mathrm{R}^{2\vdot}\right]-k_{d\rightarrow s}^F\left[\mathrm{R}\right]\\
        \dv{t}\left[\mathrm{R}^{2\vdot}\right]=&k_{d\rightarrow s}^F\left[\mathrm{R}\right]-k_{d\rightarrow s}^B\left[\mathrm{R}^{2\vdot}\right]\\
        &\Rightarrow k_{d\rightarrow s}^B\left[\mathrm{R}^{2\vdot}\right]=k_{d\rightarrow s}^F\left[\mathrm{R}\right]
    \end{aligned}
\end{equation}

Accounting for the fact that on a bare diamond surface, $\left[\mathrm{R}\right]+\left[\mathrm{R}^{2\vdot}\right]=1$, we find:
\begin{equation}
    \begin{aligned}
        &k_{d\rightarrow s}^B\left[\mathrm{R}^{2\vdot}\right]=k_{d\rightarrow s}^F\left(1-\left[\mathrm{R}^{2\vdot}\right]\right)\\
        &\Rightarrow
            \left[\mathrm{R}^{2\vdot}\right]=\frac{k_{d\rightarrow s}^F}{k_{d\rightarrow s}^B}\frac{1}{1+\frac{k_{d\rightarrow s}^F}{k_{d\rightarrow s}^B}}
    \end{aligned}
\end{equation}

Unfortunately, our approach do not allow for determination of the TS lying between $\mathrm{R}^{2\vdot}$ and $\mathrm{R}$, which is required to compute either $k_{d\rightarrow s}^F$ or $k_{d\rightarrow s}^B$. However, in the equation above, only the ratio between these two rates is required. Following the VTST methodology to calculate the reaction rate coefficient, we can express this ratio without resorting to any of the properties of the TS:

\begin{equation}
    \frac{k_{d\rightarrow s}^F}{k_{d\rightarrow s}^B}=\frac{Q_F}{Q_I}\exp\left(-\frac{\Delta E_{ZPE}}{k_BT}\right)\times\exp\left(-\frac{\Delta E}{k_BT}\right)
\end{equation}

where $\Delta E_{ZPE}$ is the difference between zero-point energies of final and initial states. At T=1200 K, this ratio is $7.766\times10^{-5}$, leading to the following distribution between $\mathrm{R}$ and $\mathrm{R}^{2\vdot}$:
\begin{equation}
    \left[\mathrm{R}\right]=1-\left[\mathrm{R}^{2\vdot}\right]\text{ $ $ with $ $ }\left[\mathrm{R^{2\vdot}}\right]=\text{7.765$\times10^{-5}$}
\end{equation}

This result should not come as a surprise when looking at the $\Delta E_n$ values of reaction 5 and 6 in Table \ref{tab:reaction_1-6}: the difference in energy between the single- and double-like dimers is more than 1 eV, strongly favouring double-bonded over single-bonded ones, in agreement with previous work.\cite{Hukka1994} In that respect, we assume in the following that the steady-state concentration of $\mathrm{R}^{2\vdot}$ is negligible (on an isolated surface), and that under other conditions (\textit{e.g.} hydrogen atmosphere containing radicals), single- and double-bonded dimers can be considered under a unique label $\mathrm{R}$.

\section{VTST}
In the main paper, we present results obtained through variational transition state theory (VTST). The associated reactions have in common that they do not possess an energy barrier (\textit{c.f.} Fig \ref{fig:Energy_Profile}). 

To dive into the technical detail of the VTST calculation presented in this work, we remind that this theory demands a sampling of the reaction path. For a given temperature, we compute the reaction rate coefficient of each point. The loose transition state (TS) is then identified as the one resulting in the lowest reaction rate coefficient, therefore representing the kinetic bottleneck of the reaction for this temperature. The first reaction shown in the main paper ($n$=1, $\mathrm{RH}\leftrightharpoons\mathrm{R}^{\vdot}+\mathrm{H}^{\vdot}$, whose energy profile is shown in Fig. \ref{fig:Energy_Profile}) concerns the unimolecular dissociation/bimolecular recombination of a hydrogen radical and a half-passivated dimer. Its reaction rate coefficient as a function of temperature is shown in Fig. \ref{fig:VTST_fit}. It is worth mentioning that not all sampling points are relevant: only the ones highlighted in red in Fig. \ref{fig:Energy_Profile} are actually involved in the reaction rate coefficient calculation. 

In Fig. 3, we report for all four barrierless reactions (1,3,3$'$ and 4) the fluxes $k\left(T,\bar{s}\right)$ along with the corresponding plot highlighting the reaction rate coefficient $k(T)$.

In Fig. \ref{fig:VTST_fit}, the reaction rate coefficient (blue curve) has been fitted (red curve) to the following equation:
\begin{equation}
    k_n\left(T\right)=a_nT^b_n\exp\left(\frac{-c_n}{k_BT}\right).
    \label{eq:fit_param}
\end{equation}

To allow the readers to reproduce best the reaction rate coefficients of all the reaction shown in this work, we provide them with the three parameters of equation \ref{eq:fit_param} in Table \ref{tab:fit_parameters}. To also inform about the quality of the fit, we provide the mean relative error $\mu$ (that is the mean relative difference between the calculation shown in the main paper and the fit) along with its standard deviation $\sigma$. An example can be seen in Fig \ref{fig:mu_sigma_example}.

\begin{figure}[!h]
    \subfloat[]{\includesvg[scale=0.14]{Figures/ENCUT_KPTS.svg}}
    \hfill
    \subfloat[]{\includesvg[scale=0.14]{Figures/dv_ENCUT_KPTS.svg}}
    \hfill
    \subfloat[]{\includesvg[scale=0.14]{Figures/sigma_N_z.svg}}
    \hfill
    \subfloat[]{\includesvg[scale=0.14]{Figures/dv_sigma_N_z.svg}}
    \caption{\label{fig:conv_test}(a): Evolution of the total energy of the reconstructed surface unit cell (2$\times$2, 20 carbon layers plus passivating hydrogen atoms on top and bottom of the slab and a 12 \AA $ $ distance between periodic copies) with respect to the cut-off energy of the plane-wave basis set (blue line) and to the number of k-points considered along the $a$ and $b$ directions of the reciprocal lattice (orange line) on a linear-scaled graph. For the former, a 7$\times$7$\times$1 k-point sampling was used while the latter used a kinetic cut-off energy of 600 eV. (b): Derivative of (a) on a logarithm-scaled graph. For 600 eV or more, results shows that variations due to the cut-off energy account for less than 0.01 meV. The same reasoning applies to a k-point sampling of the Brillouin zone greater than 7$\times$7$\times$1. (c): Evolution of the surface energy of the reconstructed surface unit cell (2$\times$2) with respect to the number of layers $N$ (blue line) and to the distance between periodic copies of the slab along the $z$-axis (orange line) on a linear-scaled graph. The former convergence study have been performed with a fixed distance of 12 \AA $ $ between periodic copies of the slab while the latter have been performed with a fixed number of 9 layers. For both convergence study calculations were performed with a 600 eV kinetic cut-off energy and a 6$\times$6$\times$1 k-point sampling. (d): Derivative of (c). Eleven or more layers, along with a distance greater than 8 \AA $ $ between periodic copies of the slab, were found to be sufficient to ensure that variations due to these parameters account for less than 0.1 meV.}
\end{figure}

\begin{figure}[!htbp]
    \centering
        \includegraphics[scale=0.23]{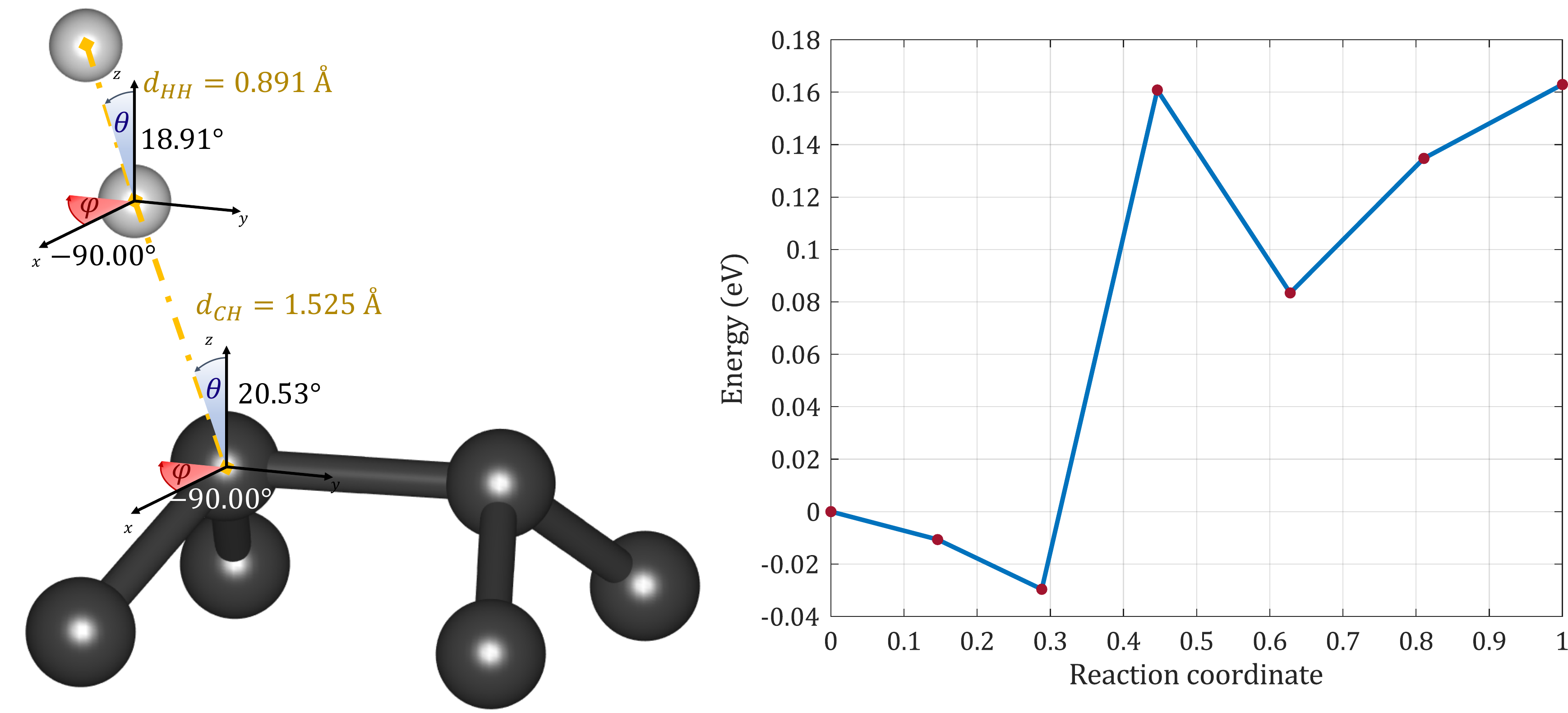}
    \caption{Hydrogen radical attack on a half-passivated dimer (reaction 6 in Table \ref{tab:reaction_1-6}): $\mathrm{R^{\vdot}H}+\mathrm{H}^{\vdot}\leftrightharpoons\mathrm{R}^{2\vdot}+\mathrm{H}_2$. Left: Geometry of the tTS resulting in a single-bonded dimer. Right: Energy of the images of the hydrogen radical attack on the remaining hydrogen on the dimer.}
    \label{fig:SI_reaction_4_prim}
\end{figure}

\begin{figure}[!htbp]
    \subfloat[\label{fig:sub1A}]{\includesvg[scale=0.28]{Figures/LTS_map_001.svg}}
    \subfloat[\label{fig:sub1B}]{\includesvg[scale=0.28]{Figures/LTS_rate_001.svg}}

    \subfloat[\label{fig:sub3A}]{\includesvg[scale=0.28]{Figures/LTS_map_003.svg}}
    \subfloat[\label{fig:sub3B}]{\includesvg[scale=0.28]{Figures/LTS_rate_003.svg}}
\end{figure}

\begin{figure}[!htp]
    \ContinuedFloat
    \subfloat[\label{fig:sub4A}]{\includesvg[scale=0.28]{Figures/LTS_map_004.svg}}
    \subfloat[\label{fig:sub4B}]{\includesvg[scale=0.28]{Figures/LTS_rate_004.svg}}

    \subfloat[\label{fig:sub5A}]{\includesvg[scale=0.28]{Figures/LTS_map_005.svg}}
    \subfloat[\label{fig:sub5B}]{\includesvg[scale=0.28]{Figures/LTS_rate_005.svg}}
    \caption{VTST applied to the backward reaction of $n$=1,3,3$'$ and the forward reaction of $n$=4. Left-column maps (a,c,e,g) show the reaction rate coefficient as a function of temperature and reaction coordinate. Purple lines show how the loose TS changes with temperature. Right-column graphs (b,d,f,h) show the contribution of the relevant sampling point (\textit{e.g.} red points in Fig. \ref{fig:Energy_Profile}) to the reaction rate coefficients.}
\end{figure}

\begin{figure}[!htp]
    \subfloat[\label{fig:Energy_Profile}]{\includegraphics[scale=0.25]{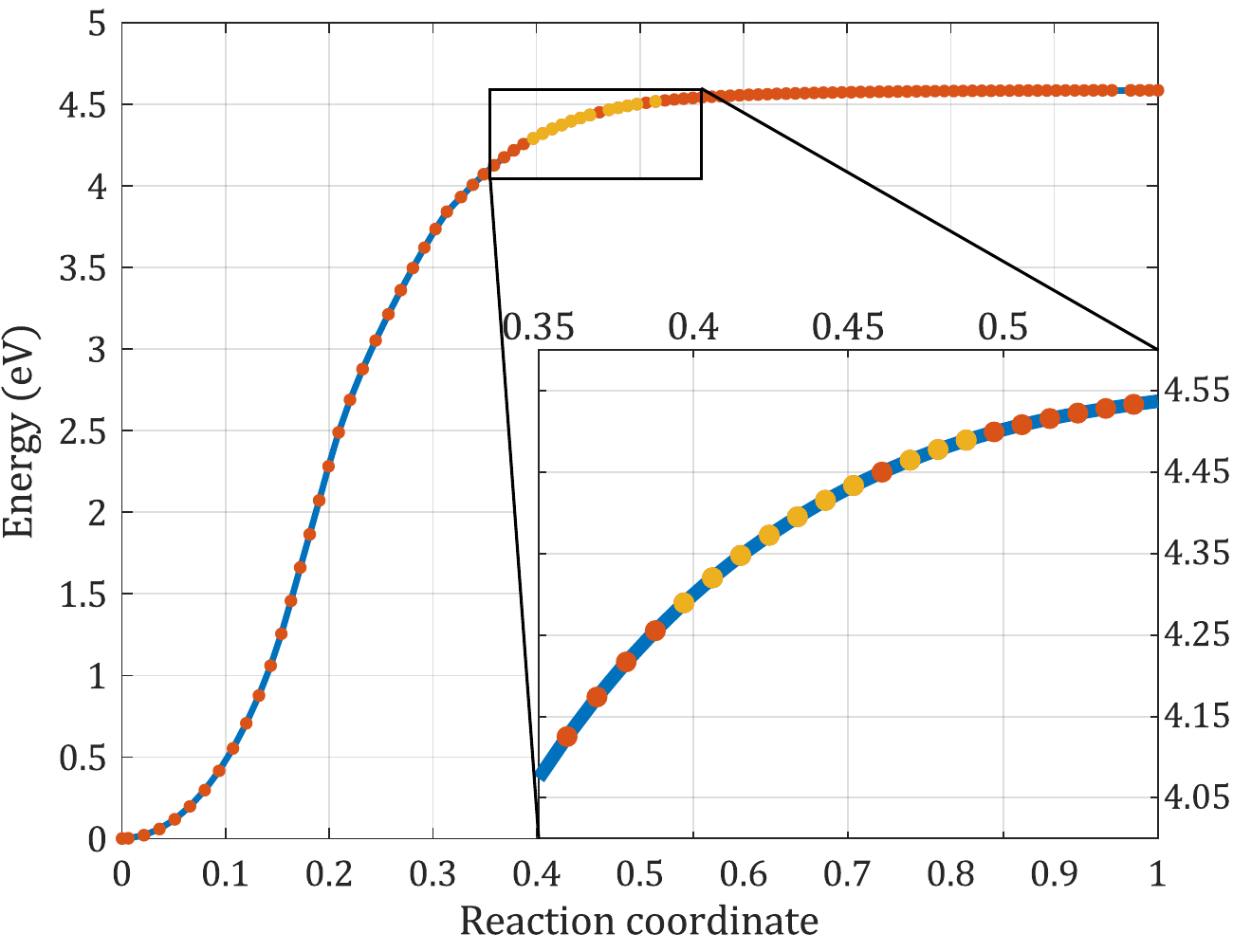}}
    \hfill
    \subfloat[\label{fig:VTST_fit}]{\includesvg[scale=0.40]{Figures/VTST_Fit.svg}}
    \hfill
    \subfloat[\label{fig:mu_sigma_example}]{\includesvg[scale=0.40]{Figures/mu_sigma_example_bw.svg}}
    \caption{(a): Energy profile of the reaction $\mathrm{RH}\leftrightharpoons\mathrm{R}^{\vdot}+\mathrm{H}^{\vdot}$. The sampling is uniform over the reaction path, and the relevant sampling area is highlighted in yellow. (b): reaction rate coefficient of the reaction $\mathrm{RH}\leftrightharpoons\mathrm{R}^{\vdot}+\mathrm{H}^{\vdot}$ obtained through VTST calculation (blue curve) and fit of the reaction rate coefficient (red curve). (c): Relative error between the VTST reaction rate coefficient and the fit.}
\end{figure}

\begin{table*}[!htbp]
    \centering
        \caption{Summary of the fitting parameters of the reactions presented in the main paper. $\mathcal{T}$ indicate a tight TS and $\mathcal{L}$ a loose TS. $b$ is unitless, whereas $a$ is expressed in units of s$^{-1}$ (\dag) or cm$^{3}\cdot$mol$^{-1}\cdot$s$^{-1}$ (\ddag), depending on the absence or presence of a gas reactant. Numbers in square brackets denote power of 10. Unless otherwise mentioned, fitting analysis is performed over temperatures ranging from 500 to 2500 K.}
        \label{tab:reaction_1-6}
        \begin{tabularx}{\textwidth}{c @{\extracolsep{\fill}}r@{\extracolsep{\fill}}c@{\extracolsep{\fill}}l@{\extracolsep{\fill}}c@{\extracolsep{\fill}}c@{\extracolsep{\fill}}c@{\extracolsep{\fill}}c@{\extracolsep{\fill}}c@{\extracolsep{\fill}}c@{\extracolsep{\fill}}c@{\extracolsep{\fill}}c@{\extracolsep{\fill}}c}
            \hline
            \multicolumn{13}{|c|}{Hydrogen ad/desorption reactions}\\
            \hline
            \hline
            \multirow{2}{*}{n} & \multicolumn{3}{c}{\multirow{2}{*}{Reaction}} & \multirow{2}{*}{} & \multirow{2}{*}{} & \multirow{2}{*}{} & \multirow{2}{*}{$a_n$} & \multirow{2}{*}{$b_n$} & \multirow{2}{*}{$c_n$ (eV)} & \multirow{2}{*}{$\mu$ (\%)} & \multirow{2}{*}{$\sigma$ (\%)} & \multirow{2}{*}{T (K)}\\ \\
            \hline
            \multirow{2}{*}{ 1} & \multirow{2}{*}{$\mathrm{RH}_2$} & \multirow{2}{*}{$\leftrightharpoons$} 
            & \multirow{2}{*}{$\mathrm{R^{\vdot}H}+\mathrm{H}^{\vdot}$} & \multirow{2}{*}{$\mathcal{L}$}
                    & Fwd &  \dag & 2.384[+15] & 0.229 &    4.3767 & 0.59 & 0.43 & \multirow{2}{*}{[860;2500]}\\
            & & & & & Bwd & \ddag & 3.059[+11] & 1.048 & $-$0.0136 & 0.35 & 0.23 &\\
            \multirow{2}{*}{ 2} & \multirow{2}{*}{$\mathrm{RH}_2+\mathrm{H}^{\vdot}$} & \multirow{2}{*}{$\leftrightharpoons$} 
            & \multirow{2}{*}{$\mathrm{R^{\vdot}H}+\mathrm{H}_2$} & \multirow{2}{*}{$\mathcal{T}$}
                    & Fwd & \ddag & 4.526[+07] & 1.863 &    0.0004 & 0.42 & 0.26 & \multirow{2}{*}{$-$}\\
            & & & & & Bwd & \ddag & 5.435[+03] & 2.702 &    0.1962 & 0.45 & 0.35 &\\
            \multirow{2}{*}{ 3} & \multirow{2}{*}{$\mathrm{R^{\vdot}H}$} & \multirow{2}{*}{$\leftrightharpoons$} 
            & \multirow{2}{*}{$\mathrm{R}+\mathrm{H}^{\vdot}$} & \multirow{2}{*}{$\mathcal{L}$}
                    & Fwd &  \dag & 5.027[+13] & 0.564 &    3.3782 & 0.63 & 0.48 & \multirow{2}{*}{$-$}\\
            & & & & & Bwd & \ddag & 4.626[+10] & 1.137 & $-$0.0582 & 1.61 & 0.87 &\\
            \multirow{2}{*}{ 3$'$} & \multirow{2}{*}{$\mathrm{R^{\vdot}H}$} & \multirow{2}{*}{$\leftrightharpoons$} 
            & \multirow{2}{*}{$\mathrm{R}^{2\vdot}+\mathrm{H}^{\vdot}$} & \multirow{2}{*}{$\mathcal{L}$}
                    & Fwd &  \dag & 2.100[+15] & 0.274 &    4.4672 & 0.68 & 0.45 & \multirow{2}{*}{[800;2500]}\\
            & & & & & Bwd & \ddag & 3.719[+11] & 1.014 & $-$0.0181 & 0.85 & 0.58 &\\
            \multirow{2}{*}{ 4} & \multirow{2}{*}{$\mathrm{R^{\vdot}H}+\mathrm{H}^{\vdot}$} & \multirow{2}{*}{$\leftrightharpoons$} 
            & \multirow{2}{*}{$\mathrm{R}+\mathrm{H}_2$} & \multirow{2}{*}{$\mathcal{L}$}
                    & Fwd & \ddag & 3.308[+06] & 2.203 & $-$0.2055 & 2.42 & 2.37 & \multirow{2}{*}{[610;2500]}\\
            & & & & & Bwd & \ddag & 1.208[+02] & 3.167 &    0.898 & 2.82 & 2.67 &\\
            \multirow{2}{*}{ 4$'$} & \multirow{2}{*}{$\mathrm{R^{\vdot}H}+\mathrm{H}^{\vdot}$} & \multirow{2}{*}{$\leftrightharpoons$} 
            & \multirow{2}{*}{$\mathrm{R}^{2\vdot}+\mathrm{H}_2$} & \multirow{2}{*}{$\mathcal{T}$}
                    & Fwd & \ddag & 6.612[+07] & 1.861 &    0.0765 & 0.43 & 0.27 & \multirow{2}{*}{$-$}\\
            & & & & & Bwd & \ddag & 1.176[+04] & 2.614 &    0.1872 & 0.57 & 0.43 &\\
            \hline
            \multicolumn{13}{|c|}{Migration of a hydrogen atom on a clean (100) surface}\\
            \hline
            \hline
            \multirow{2}{*}{n} & \multicolumn{3}{c}{\multirow{2}{*}{Reaction}} & \multirow{2}{*}{} & \multirow{2}{*}{} & \multirow{2}{*}{} & \multirow{2}{*}{$a$} & \multirow{2}{*}{$b$} & \multirow{2}{*}{$c$ (eV)} & \multirow{2}{*}{$\mu$ (\%)} & \multirow{2}{*}{$\sigma$ (\%)} & \multirow{2}{*}{T (K)}\\ \\
            \hline
            \multirow{2}{*}{ 5} & \multirow{2}{*}{$\mathrm{R}_{A}^{\vdot}\mathrm{H}+\mathrm{R}_{B}$} & \multirow{2}{*}{$\leftrightharpoons$} & \multirow{2}{*}{$\mathrm{R}_{A}+\mathrm{R}_{B}^{\vdot}\mathrm{H}$} & \multirow{2}{*}{$\mathcal{T}$}
                    & \multirow{2}{*}{Sym.} & \multirow{2}{*}{\dag} & \multirow{2}{*}{7.472[+01]} & \multirow{2}{*}{3.346} & \multirow{2}{*}{2.390} & \multirow{2}{*}{11.62} & \multirow{2}{*}{14.27} & \multirow{2}{*}{$-$}\\ \\
            \multirow{2}{*}{ 6} & \multirow{2}{*}{$\mathrm{R}_{B}^{\vdot}\mathrm{H}+\mathrm{R}_{C}$} & \multirow{2}{*}{$\leftrightharpoons$} & \multirow{2}{*}{$\mathrm{R}_{B}+\mathrm{R}_{C}^{\vdot}\mathrm{H}$} & \multirow{2}{*}{$\mathcal{T}$}
                    & \multirow{2}{*}{Sym.} & \multirow{2}{*}{\dag} & \multirow{2}{*}{9.165[+06]} & \multirow{2}{*}{1.991} & \multirow{2}{*}{2.651} & \multirow{2}{*}{4.69} & \multirow{2}{*}{3.79} & \multirow{2}{*}{$-$}\\ \\
            \multirow{2}{*}{ 7} & \multirow{2}{*}{$\mathrm{R}_{C}^{\vdot}\mathrm{H}$} & \multirow{2}{*}{$\leftrightharpoons$} & \multirow{2}{*}{$\mathrm{R}_{D}^{\vdot}\mathrm{H}$} & \multirow{2}{*}{$\mathcal{T}$}
                    & \multirow{2}{*}{Sym.} & \multirow{2}{*}{\dag} & \multirow{2}{*}{9.756[+06]} & \multirow{2}{*}{1.823} & \multirow{2}{*}{1.938} & \multirow{2}{*}{5.31} & \multirow{2}{*}{4.67} & \multirow{2}{*}{$-$}\\ \\
            \hline
            \multicolumn{13}{|c|}{Migration of a single hydrogen vacancy on a H-passivated (100) surface}\\
            \hline
            \hline
            \multirow{2}{*}{n} & \multicolumn{3}{c}{\multirow{2}{*}{Reaction}} & \multirow{2}{*}{} & \multirow{2}{*}{} & \multirow{2}{*}{} & \multirow{2}{*}{$a$} & \multirow{2}{*}{$b$} & \multirow{2}{*}{$c$ (eV)} & \multirow{2}{*}{$\mu$ (\%)} & \multirow{2}{*}{$\sigma$ (\%)} & \multirow{2}{*}{T (K)}\\ \\
            \hline
            \multirow{2}{*}{8} & \multirow{2}{*}{$\mathrm{R}_{A}^{\vdot}\mathrm{H}+\mathrm{R}_{B}\mathrm{H}_2$} & \multirow{2}{*}{$\leftrightharpoons$} & \multirow{2}{*}{$\mathrm{R}_{A}\mathrm{H}_2+\mathrm{R}_{B}^{\vdot}\mathrm{H}$} 
 & \multirow{2}{*}{$\mathcal{T}$}
                    & \multirow{2}{*}{Sym.} & \multirow{2}{*}{\dag} & \multirow{2}{*}{9.139[-01]} & \multirow{2}{*}{3.902} & \multirow{2}{*}{2.386} & \multirow{2}{*}{13.97} & \multirow{2}{*}{19.22} & \multirow{2}{*}{$-$}\\ \\
            \multirow{2}{*}{9} & \multirow{2}{*}{$\mathrm{R}_{B}^{\vdot}\mathrm{H}+\mathrm{R}_{C}\mathrm{H}_2$} & \multirow{2}{*}{$\leftrightharpoons$} & \multirow{2}{*}{$\mathrm{R}_{B}\mathrm{H}_2+\mathrm{R}_{C}^{\vdot}\mathrm{H}$} 
 & \multirow{2}{*}{$\mathcal{T}$}
                    & \multirow{2}{*}{Sym.} & \multirow{2}{*}{\dag} & \multirow{2}{*}{9.672[+06]} & \multirow{2}{*}{2.022} & \multirow{2}{*}{2.249} & \multirow{2}{*}{4.90} & \multirow{2}{*}{3.96} & \multirow{2}{*}{$-$}\\ \\
            \multirow{2}{*}{10} & \multirow{2}{*}{$\mathrm{R}_{C}^{\vdot}\mathrm{H}$} & \multirow{2}{*}{$\leftrightharpoons$} & \multirow{2}{*}{$\mathrm{R}_{D}^{\vdot}\mathrm{H}$} 
 & \multirow{2}{*}{$\mathcal{T}$}
                    & \multirow{2}{*}{Sym.} & \multirow{2}{*}{\dag} & \multirow{2}{*}{1.004[+07]} & \multirow{2}{*}{1.886} & \multirow{2}{*}{1.878} & \multirow{2}{*}{5.23} & \multirow{2}{*}{4.48} & \multirow{2}{*}{$-$}\\ \\
            \hline
            \multicolumn{13}{|c|}{Migration of a double hydrogen vacancy on a H-passivated (100) surface}\\
            \hline
            \hline
            \multirow{2}{*}{n} & \multicolumn{3}{c}{\multirow{2}{*}{Reaction}} & \multirow{2}{*}{} & \multirow{2}{*}{} & \multirow{2}{*}{} & \multirow{2}{*}{$a$} & \multirow{2}{*}{$b$} & \multirow{2}{*}{$c$ (eV)} & \multirow{2}{*}{$\mu$ (\%)} & \multirow{2}{*}{$\sigma$ (\%)} & \multirow{2}{*}{T (K)}\\ \\
            \hline
            \multirow{2}{*}{11} & \multirow{2}{*}{$\mathrm{R}_{\alpha}^{\vdot}\mathrm{H}+\mathrm{R}_{A}^{\vdot}\mathrm{H}+\mathrm{R}_{B}\mathrm{H}_2 $} & \multirow{2}{*}{$\leftrightharpoons$} & \multirow{2}{*}{$\mathrm{R}_{\alpha}^{\vdot}\mathrm{H}+\mathrm{R}_{A}\mathrm{H}_2+\mathrm{R}_{B}^{\vdot}\mathrm{H}$}
 & \multirow{2}{*}{$\mathcal{T}$}
                    & Fwd &  \dag & 3.512[+01] & 3.472 & 2.577 & 12.29 & 15.62 & \multirow{2}{*}{$-$}\\
            & & & & & Bwd &  \dag & 3.355[+01] & 3.474 & 2.602 & 12.31 & 15.63 &\\
            \multirow{2}{*}{12} & \multirow{2}{*}{$\mathrm{R}_{\alpha}^{\vdot}\mathrm{H}+\mathrm{R}_{B}^{\vdot}\mathrm{H}+\mathrm{R}_{C}\mathrm{H}_2 $} & \multirow{2}{*}{$\leftrightharpoons$} & \multirow{2}{*}{$\mathrm{R}_{\alpha}^{\vdot}\mathrm{H}+\mathrm{R}_{B}\mathrm{H}_2+\mathrm{R}_{C}^{\vdot}\mathrm{H}$}
 & \multirow{2}{*}{$\mathcal{T}$}
                    & Fwd &  \dag & 3.268[+07] & 1.847 & 2.353 & 4.47 & 3.63 & \multirow{2}{*}{$-$}\\
            & & & & & Bwd &  \dag & 3.279[+07] & 1.849 & 2.342 & 4.48 & 3.64 &\\
            \multirow{2}{*}{13} & \multirow{2}{*}{$\mathrm{R}_{\alpha}^{\vdot}\mathrm{H}+\mathrm{R}_{C}^{\vdot}\mathrm{H} $} & \multirow{2}{*}{$\leftrightharpoons$} & \multirow{2}{*}{$\mathrm{R}_{\alpha}^{\vdot}\mathrm{H}+\mathrm{R}_{D}^{\vdot}\mathrm{H}$}
 & \multirow{2}{*}{$\mathcal{T}$}
                    & Fwd &  \dag & 5.806[+07] & 1.614 & 2.259 & 4.50 & 3.87 & \multirow{2}{*}{$-$}\\
            & & & & & Bwd &  \dag & 5.810[+07] & 1.611 & 2.276 & 4.49 & 3.87 &\\
            \multirow{2}{*}{14} & \multirow{2}{*}{$\mathrm{R}^{\vdot}_{\alpha}\mathrm{H}+\mathrm{R}_{D}^{\vdot}\mathrm{H} $} & \multirow{2}{*}{$\leftrightharpoons$} & \multirow{2}{*}{$\mathrm{R}_{\alpha,E}+\mathrm{R}_D\mathrm{H}_2$}
 & \multirow{2}{*}{$\mathcal{T}$}
                    & Fwd &  \dag & 1.414[+03] & 2.698 & 1.849 & 9.20 & 8.28 & \multirow{2}{*}{$-$}\\
            & & & & & Bwd &  \dag & 2.774[+04] & 2.978 & 2.719 & 9.93 & 8.90 &\\
            \multirow{2}{*}{15} & \multirow{2}{*}{$\mathrm{R}_{\alpha,E}+\mathrm{R}_F\mathrm{H}_2$} & \multirow{2}{*}{$\leftrightharpoons$} & \multirow{2}{*}{$\mathrm{R}^{\vdot}_{\alpha}\mathrm{H}+\mathrm{R}^{\vdot}_{F}\mathrm{H}$}
 & \multirow{2}{*}{$\mathcal{T}$}
                    & Fwd &  \dag & 5.510[+06] & 2.110 & 3.019 & 5.02 & 4.02 & \multirow{2}{*}{$-$}\\
            & & & & & Bwd &  \dag & 1.742[+07] & 1.888 & 2.143 & 4.66 & 3.82 &\\
            \hline
        \end{tabularx}
        \label{tab:fit_parameters}
\end{table*}

\newpage 
$ $

\newpage 
$ $

\bibliography{Biblio}